\documentclass[referee]{raa}            

\usepackage{graphicx,times}             
\usepackage{natbib}
\usepackage{amssymb,amsmath}
\bibpunct{(}{)}{;}{a}{}{,}
\usepackage{siunitx}
\usepackage{sidecap}
\usepackage{enumitem}

\usepackage[pagebackref=true]{hyperref}

\begin{document}

  \title{Early Near-Infrared Excess and Rapid Disk-Corona Evolution in the Tidal Disruption Event 2024aepd
}

   \volnopage{Vol.0 (20xx) No.0, 000--000}      
   \setcounter{page}{1}          

   \author{Yongxin Wu 
      \inst{1,2}
   \and Yanan Wang
      \inst{1}
   \and Thomas M. Reynolds
      \inst{3,4,5}
   \and Shuyuan Wei
      \inst{2}
   \and Shiyan Zhong
      \inst{6}
   \and Zikun Lin
      \inst{7}
   \and Megan Newsome
      \inst{8}
   \and Sebastian Gomez
      \inst{8}
   \and Iair Arcavi
      \inst{9}
   \and Panos Charalampopoulos
      \inst{10,11}
   \and Chun Chen
      \inst{12,13}
   \and Rongfeng Shen
      \inst{12,13}
   \and Ning-Chen Sun
      \inst{2,1,14}      
   \and David Aguado
      \inst{15,16}
   \and Ismael P\'erez-Fournon
      \inst{15,16}
   \and Fr\'ed\'erick Poidevin
      \inst{15,16}
   \and Zhongnan Dong
      \inst{1,10}      
   \and Niu Li
      \inst{1}
   \and Weijian Guo
      \inst{1}
   \and Hu Zou
      \inst{1}
   \and Jingbo Sun
      \inst{17}
   \and Nieves Castro-Rodr\'iguez
      \inst{14,18}
   \and Antonio Cabrera-Lavers
      \inst{14,18}
   \and Ning Jiang
      \inst{19,20}
   \and Hengxiao Guo
      \inst{17}
   \and J.~P. Anderson 
      \inst{21}
   \and  Tom\'as E. M\"uller-Bravo 
      \inst{22,23}
   \and Seppo Mattila
      \inst{3,24}
   \and Claudia P. Guti\'errez 
      \inst{10,25}
   \and Amit Kumar
      \inst{26}
   \and G. Pignata
      \inst{27}
   \and Xiangkun Liu
      \inst{6,28}
   \and R. Dastidar
      \inst{29}
   \and Brajesh Kumar
      \inst{6,28}
   \and Xiaowei Liu
      \inst{6} 
    \and Bin Ma
      \inst{12,13} 
    \and M. Dennefeld
      \inst{30}
    \and Francesca Onori
      \inst{31}
    \and Lydia Makrygianni
      \inst{32}
    \and Mariusz Gromadzki
      \inst{33}
    \and Xuan Fang
      \inst{1,2,34,35}
   }


   \institute{Key Laboratory of Optical Astronomy, National Astronomical Observatories, Chinese Academy of Sciences, Beijing 100101, People’s Republic of China; {\it wangyn@bao.ac.cn}\\
        \and
             School of Astronomy and Space Sciences, University of Chinese Academy of Sciences, Beijing 100049, People’s Republic of China\\
        \and
             Tuorla observatory, Department of Physics and Astronomy, University of Turku, FI-20014 Turku, Finland\\
        \and 
             Cosmic Dawn Center (DAWN)\\
        \and
             Niels Bohr Institute, University of Copenhagen, Jagtvej 128, DK-2200, Copenhagen N, Denmark\\
        \and
             South-Western Institute for Astronomy Research, Yunnan University, Kunming, 650500 Yunnan, People’s Republic of China\\
        \and
             Department of Astronomy, Xiamen University, Xiamen, Fujian 361005, People’s Republic of China\\
        \and
             Department of Astronomy, The University of Texas at Austin, 2515 Speedway, Stop C1400, Austin, TX 78712, USA\\
        \and
             The School of Physics and Astronomy, Tel Aviv University, Tel Aviv 69978, Israel\\
        \and
             Institute of Space Sciences (ICE-CSIC), Campus UAB, Carrer de Can Magrans, s/n, E-08193 Barcelona, Spain\\
        \and
             Finnish Centre for Astronomy with ESO (FINCA), FI-20014 University of Turku, Finland\\
        \and
             School of Physics and Astronomy, Sun Yat-Sen University, Zhuhai 519082, China\\
        \and
             CSST Science Center for the Guangdong-Hong Kong-Macau Greater Bay Area, Sun Yat-Sen University, Zhuhai 519082, China\\
        \and Institute for Frontiers in Astronomy and Astrophysics, Beijing Normal University, Beijing 102206, People’s Republic of China\\
        \and
             Instituto de Astrof\'isica de Canarias, C/V\'ia L\'actea, s/n, E-38205 La Laguna, Tenerife, Spain\\
        \and
             Departamento de Astrof\'isica, Universidad de La Laguna (ULL), E-38206, LaLaguna, Tenerife, Spain\\
        \and
             Shanghai Astronomical Observatory, Chinese Academy of Sciences, 80 Nandan Road, Shanghai 200030, People’s Republic of China\\
        \and 
             GRANTECAN, Cuesta de San Jos\'e s/n, 38712 Bre\~{n}a Baja, La Palma, Spain\\
        \and 
             CAS Key laboratory for Research in Galaxies and Cosmology, Department of Astronomy, University of Science and Technology of China, Hefei, 230026, China\\
        \and 
             School of Astronomy and Space Sciences, University of Science and Technology of China, Hefei, 230026, China\\
        \and 
             European Southern Observatory, Alonso de C\'ordova 3107, Vitacura, Casilla 19001, Santiago, Chile\\
        \and 
             School of Physics, Trinity College Dublin, The University of Dublin, Dublin 2, Ireland\\
        \and
             Instituto de Ciencias Exactas y Naturales (ICEN), Universidad Arturo Prat, Chile\\
        \and
             School of Sciences, European University Cyprus, Diogenes Street, Engomi, 1516 Nicosia, Cyprus\\
        \and 
              Institut d'Estudis Espacials de Catalunya (IEEC), Edifici RDIT, Campus UPC, 08860 Castelldefels (Barcelona), Spain\\
        \and 
             Centre for Electronic Imaging, School of Physical Sciences, The Open University, Walton Hall, Milton Keynes MK7 6AA, UK\\
        \and
             Instituto de Alta Investigaci\'on, Universidad de Tarapac\'a, Casilla 7D, Arica, Chile\\
        \and
             Yunnan Key Laboratory of Survey Science, Yunnan University, Kunming, Yunnan 650500, People's Republic of China\\
        \and
             Istituto Nazionale di Astrofisica, Osservatorio Astronomico di Brera, via E. Bianchi 46, 23807 Merate (LC), Italy\\
        \and
             Institut d’Astrophysique de Paris (IAP), Sorbonne Universit\'e, CNRS, Paris, France\\
        \and
             INAF - Osservatorio Astronomico di Roma, Via Frascati, 33, 00078 Monte Porzio Catone (RM), Italy\\
        \and 
             Department of Physics, Lancaster University, Lancaster LA1 4YB, UK\\
        \and
             Astronomical Observatory, University of Warsaw, Al. Ujazdowskie 4, 00-478 Warszawa, Poland\\
        \and
             Xinjiang Astronomical Observatory, Chinese Academy of Sciences, 150 Science 1-Street, Urumqi, Xinjiang 830011, People’s Republic of China\\
        \and 
             Laboratory for Space Research, Faculty of Science, The University of Hong Kong, Pokfulam Road, Hong Kong, People’s Republic of China\\
\vs\no
   {\small Received 20xx month day; accepted 20xx month day}}

\abstract{We present multi-wavelength observations of the tidal disruption event (TDE) 2024aepd, spanning primarily the first $\sim$300~days after discovery. 
A prominent near-infrared (NIR) excess is detected as early as $\sim$40~days. Its nearly flat power-law spectrum strongly deviates from the Rayleigh--Jeans tail of the UV--optical blackbody. Although a conventional dust-echo origin cannot be completely ruled out, free--free emission from a reprocessing photospheric envelope provides a more plausible explanation.
The spectral break between the UV--optical and NIR components shifts to higher frequencies, while the inferred density-profile index remains nearly constant, suggesting evolving reprocessing conditions within a broadly unchanged density structure. In addition, the X-ray spectrum is initially dominated by a thermal disk component accompanied by a hard excess. From $\sim$178~days onward, the spectrum becomes power-law dominated and subsequently hardens, indicating the rapid emergence and strengthening of a hot corona. These results provide evidence for frequency-dependent reprocessing at early times and for the rapid development of a disk--corona system, placing new constraints on the structure and evolution of the reprocessing layer around supermassive black holes. 
}

   \authorrunning{}            
   \titlerunning{NIR excess and disk-corona evolution in TDE 2024aepd}  

   \maketitle

%
%
\section{Introduction}           
\label{sect:intro}
Tidal disruption events (TDEs) occur when a star passes within the tidal radius of a supermassive black hole (SMBH), where tidal forces exceed the star’s self gravity and lead to its disruption. Approximately half of the stellar debris remains bound, returning to pericenter and circularizing through self intersecting streams before being accreted onto the SMBH \citep{1975Natur.254..295H,1988Natur.333..523R,1989ApJ...346L..13E}. This process produces luminous emission across a broad wavelength range from radio to X-rays \citep{2021ARA&A..59...21G}.

The first TDEs were discovered in the 1990s in soft X-rays by the ROentgen SATellite (ROSAT) all sky survey \citep{1996A&A...309L..35B,1999A&A...349L..45K,1999A&A...343..775K,1999A&A...350..805G}. Since then, the discovery rate of TDEs has increased rapidly (e.g., \citealt{2016MNRAS.463.3813H,2016MNRAS.455.2918H,2018ATel12263....1V,2021ApJ...908....4V,2023ApJ...942....9H}), largely owing to wide-field optical time-domain surveys such as All-Sky Automated Survey for Supernovae \citep[ASAS-SN;][]{2014ApJ...788...48S}, the
Asteroid Terrestrial-impact Last Alert System \citep[ATLAS;][]{2018PASP..130f4505T}, and the Zwicky Transient Facility \citep[ZTF;][]{2019PASP..131a8002B,2019PASP..131g8001G}. The ultraviolet (UV) and optical emission of TDEs is characterized by a blue continuum, with a blackbody temperature on the order of $10^4$~K that remains nearly constant over time, and a blackbody radius on the order of $10^{14}$~cm that steadily decreases as the TDE evolves \citep{2020SSRv..216..124V}.
The inferred radius is significantly larger than expected for a circularized debris disk. Combined with the observed diversity of TDEs, including events with and without early time X-ray detections, this suggests that more complex mechanisms are responsible for the UV and optical emission. Proposed models include reprocessing of X-ray and extreme-UV (EUV) emission by an envelope \citep{1997ApJ...489..573L,2016ApJ...827....3R,2022ApJ...937L..12M}, emission from accretion driven outflows \citep{2018ApJ...859L..20D,2022ApJ...937L..28T,2025MNRAS.539.3473Q}, collision induced outflows \citep{2020MNRAS.492..686L}, and shocks generated by self intersecting debris streams \citep{2015ApJ...806..164P,2016ApJ...830..125J,2024Natur.625..463S,2025ApJ...979..235G}. However, the relative importance of these mechanisms remains under debate, and key diagnostics needed to distinguish between these models are still lacking.

Follow-up observations of optically discovered TDEs have revealed diverse X-ray evolution, with the emission emerging either promptly or after a delay relative to the optical peak, and the subsequent light curves exhibiting either a long-term power-law decline or episodic flaring \citep{2024ApJ...966..160G}. Their spectra are typically ultrasoft, with little emission above 2~keV, and can be described by thermal blackbody models with temperatures of $10^5$ to $10^6\ \mathrm{K}$ \citep{2021SSRv..217...18S}, sometimes accompanied by a hard X-ray excess.
The thermal component is generally attributed to emission from the accretion disk \citep{1999ApJ...514..180U}, while the hard excess is commonly interpreted as thermal Comptonization, in which disk photons are up scattered by hot electrons in an optically thin corona \citep{1991ApJ...380L..51H,1995ApJ...450..876T}. Similar mechanisms are widely invoked to explain hard X-ray emission in X-ray binaries (XRBs; e.g., \citealt{2006ARA&A..44...49R}) and active galactic nuclei \citep[AGN; e.g.,][]{2007A&ARv..15....1D,2025ARA&A..63..379K}.
Some TDEs exhibit spectral state transitions from a soft, disk dominated phase to a hard, corona dominated phase on timescales of several hundred days \citep{2021ApJ...912..151W,2022ApJ...937....8Y,2025ApJ...983...29H,2026ApJ...999..265B}. These transitions are interpreted as signatures of corona formation \citep{2021ApJ...912..151W,2022ApJ...937....8Y,2024ApJ...966..160G}. In contrast to XRBs, where transitions between hard and soft states are ubiquitous during outburst evolution \citep[e.g.,][]{2006ARA&A..44...49R}, analogous behavior is rarely observed in AGN because their evolutionary timescales can extend to millions of years. TDEs therefore provide a unique laboratory for studying the formation and evolution of accretion disks and coronae around SMBHs on observable timescales.

The identification of infrared (IR) counterparts to TDEs has largely relied on observations from the Near Earth Object Wide Field Infrared Survey Explorer \citep[NEOWISE;][]{2014ApJ...792...30M}, as well as the Wide-field Infrared Survey Explorer \citep[WISE; ][]{WISE}. Observed mid-IR (MIR) excesses are typically characterized by a blackbody with a temperature of order $10^3~\mathrm{K}$, and are commonly interpreted as dust echoes, in which UV and optical radiation from the TDE is absorbed by circumnuclear dust and re-emitted in the MIR \citep{2016MNRAS.458..575L}. The MIR emission can therefore be used to estimate the bolometric luminosity of TDEs \citep{2016ApJ...829...19V}, which can be much higher than the fitted UV--optical blackbody luminosity and indicate their spectral energy distribution (SED) peak at (EUV) \citep{2025ApJ...980L..17J,2025ApJ...988L..77W}.
Recent observations have revealed a previously underexplored near-IR (NIR) excess in two early-phase TDEs, including AT2019azh \citep{2026A&A...708A.139R} and TDE~2025abcr \citep{patra2026jwstkeckobservationsoffnuclear}. This component deviates from the Rayleigh--Jeans tail of the UV--optical blackbody and has been suggested to arise from free--free reprocessing rather than dust echoes \citep{2026A&A...708A.139R}. 
In this scenario, frequency-dependent free--free absorption produces a power-law continuum shallower than the Rayleigh--Jeans slope, broadly consistent with reprocessing models \citep{2016ApJ...827....3R,2020SSRv..216..114R,2020MNRAS.492..686L}.
Similar NIR excesses have also been reported in fast blue optical transients (FBOTs), including AT2018cow \citep{2025ApJ...991..180C}, AT2024wpp \citep{2026ApJ...997L..10L}, and the FBOT-like luminous transient AT2024puz \citep{2025ApJ...995..228S}. 
Theoretically, \citet{2020MNRAS.492..686L} showed that a collision-induced outflow with $\rho\propto r^{-2}$ produces an NIR SED of $L_\nu\propto\nu^{0.5}$, while \citet{2020SSRv..216..114R} generalized the relation between the NIR spectral slope and the density profile of the envelope. \citet{2025ApJ...991..180C} further linked the outflow density and accretion rate to the outflow velocity under the same density-profile assumption. Together, these studies suggest that shallower NIR spectra may reflect shallower density profiles and may, in some cases, additionally encode information about outflow velocity and accretion conditions.
However, NIR observations of TDEs remain sparse, and current models generally treat either the density structure or the outflow dynamics in isolation. A framework incorporating both is needed for a more complete physical description. Further investigation of early-time NIR emission is therefore essential for constraining its origin and the nature of reprocessing in TDEs.

To address the above issues, multi-wavelength, long-term monitoring of TDEs is required. 
TDE~2024aepd (here after 2024aepd) is an optical transient discovered by ZTF on December 5, 2024. It is spatially coincident with the center of a quiescent galaxy at a redshift of z = 0.0835. Optical monitoring from ZTF and ATLAS, together with follow-up spectroscopy, supports the classification of this transient as a TDE\footnote{{\href{https://www.wis-tns.org/object/2024aepd}{https://www.wis-tns.org/object/2024aepd}}} \cite{Charalampopoulos2025}. 
In this paper, we present multi-wavelength observations of 2024aepd spanning radio, NIR, UV, optical, and soft X-ray bands. These data reveal an early-time NIR excess and a gradual co-evolution of the accretion disk and corona.

This paper is organized as follows. Section \ref{sect:observation} describes the observations and data reduction. In Section \ref{sect:results}, we model the multi-wavelength emission and present the physical properties and temporal evolution of the system. In Section \ref{sect:discussion}, we discuss the origin of the non-thermal X-ray emission and the NIR excess, and compare the spectral state transition and the relationship between the photon index $\Gamma$ and X-ray luminosity with those observed in other TDEs as well as in accretion-powered systems such as XRBs and AGN. Finally, we summarize our conclusions in Section \ref{sect:conclusion}.

\section{Observations and data reduction}
\label{sect:observation}
We adopt a flat $\Lambda$CDM cosmology with $H_0=67.4~\mathrm{km~s^{-1}~Mpc^{-1}}$, $\Omega_\mathrm{m}=0.315$ and $\Omega_\Lambda=0.685$ from \citet{2020A&A...641A...6P}, corresponding to a luminosity distance of 394.4~Mpc. Galactic extinction is corrected using the \citet{1999PASP..111...63F} extinction law with $R_{\mathrm{V}} = 3.1$ and $E(B-V) = 0.0411$ \citep{2011ApJ...737..103S}.
We conducted multi-wavelength observations of 2024aepd from radio to soft X-rays, primally spanning MJD~60649--60951 ($\sim$300~days).

\subsection{Optical Photometry}
\subsubsection{ZTF}
2024aepd was first discovered by ZTF on 2024 December 5 (MJD~60649), which we adopt as the reference epoch throughout this work. 
Publicly available ZTF $g$- and $r$-band photometry obtained between 2024 December 5 and 2025 March 25 was retrieved from the Lasair alert broker\footnote{\url{https://lasair-ztf.lsst.ac.uk/object/ZTF24abyhjvc/}}. The light curves were then constructed after correcting the photometry for Galactic extinction.
A total of 15 and 21 measurements in the $r$ and $g$ bands are included in our analysis, respectively.

\subsubsection{ATLAS}
ATLAS observed 2024aepd in the $c$ and $o$ bands. The photometric measurements were obtained from the ATLAS forced photometry server. To improve the signal-to-noise ratio, we stacked the forced photometry within 3~day intervals. Magnitudes and their uncertainties were calculated from the averaged fluxes and the corresponding averaged flux errors, respectively.
We include 11 stacked measurements in the $c$ band and 49 in the $o$ band in this work.

\subsubsection{Mephisto}
We obtained additional optical photometry using the 1.6~m Multi-channel Photometric Survey Telescope \citep[Mephisto;][]{2020SPIE11445E..7MY}. The monitoring began at MJD~$\sim60690$, approximately 40~days after discovery, and continued until day~$\sim160$, with a cadence of several days. Observations were primarily conducted in the $u$ (3420~\AA), $v$ (3850~\AA), $g$ (5300~\AA), $r$ (6300~\AA), $i$ (8370~\AA) and $z$ (9750~\AA) filters. Mephisto can perform simultaneous observations of a field in three optical bands ($ugi$ or $vrz$). Further details on the Mephisto filter system and data-reduction procedures are provided in \citet{2024ApJ...969..126Y}, \citet{2024ApJ...971L...2C}, and \citet{2026ApJ...997...77Z}.

We used recalibrated Gaia BP/RP (XP) spectra for photometric calibration with the synthetic photometry method. Synthetic AB magnitudes were derived by convolving XP spectra with Mephisto’s transmission curves; the photometric zero-points were determined using high-quality, non-variable stars, correcting for potential gain inaccuracies per CCD output.

All Mephisto photometry was derived by performing point-spread function (PSF) photometry on the template-subtracted images, with the final observation serving as the template. Photometric calibration was carried out using the zero-point values derived from the template image.
The Galactic extinction-corrected light curves in the $uvgr$ bands are shown in Fig.~\ref{fig1}, including 20, 13, 19, and 16 epochs, respectively. The $i$ and $z$ bands are largely excluded due to low signal-to-noise, with only 2 detections for each band respectively. The observation after the final detection yielded $5\sigma$ upper limits of 21.94 and 21.36\, mag for $i$ and $z$ band.

\subsubsection{LCO}
We subtracted from each Las Cumbres Observatory (LCO) image, using HOTPANTS \citep{Becker2015}, archival Pan-STARRS images as reference templates. We modeled the PSF of each LCO image using field stars and applied the resulting model to extract PSF photometry of the target from the difference images. The zero-points of each image were estimated by measuring the magnitudes of field stars and comparing to photometric AB magnitudes from the Pan-STARRS catalog \citep{Chambers2018}. The uncertainties of the final subtracted photometry include the combination of the photometric and zero-point determination uncertainties.

\subsubsection{Swift-UVOT}
The Neil Gehrels Swift Observatory \citep{2004ApJ...611.1005G} monitoring campaign commenced on 15 January 2025, approximately 40~days after the optical discovery, and obtained a total of 45 observations, continuing until 3 October 2025.
The target was observed by the Ultra-Violet/Optical Telescope \citep[UVOT;][]{2005SSRv..120...95R} using all or some of its six filters: $UVW2$ (1928~\AA), $UVM2$ (2246~\AA), $UVW1$ (2600~\AA), $U$ (3465~\AA), $B$ (4392~\AA), and $V$ (5468~\AA) across different UVOT observations. 

UVOT photometry was performed using the {\sc uvotsource} task with a 5" aperture centred on the optical position, while a nearby 40" radius aperture was used to estimate the background level.
We first performed photometry on individual exposures within each observation and excluded those with a magnitude of 99 and $SSS\_FACTOR = -99$, which indicate unreliable measurements. We combined all valid exposures within each observation using the {\sc uvotimsum} task and re-extracted the fluxes with {\sc uvotsource} on the combined images for each filter.

We then subtracted the host galaxy fluxes which are derived from the SED fitting of the host galaxy in these wavelength and corrected for Galactic extinction. This procedure resulted in 38 $UVW2$ observations, 32 $UVM2$, 34 $UVW1$, 36 $U$, 36 $B$, and 37 $V$ band observations. 
In addition, we find that the $B$- and $V$-band fluxes after 120~days evolve in the opposite direction to the ATLAS $c$ and $o$ measurements. Because the ATLAS photometry is based on image subtraction and is therefore less susceptible to contamination from a nearby source located 3.7\arcsec\ from the target, we exclude the $B$- and $V$-band data from this work.

\subsection{Optical Spectroscopy}
We obtained low-resolution optical spectroscopic observations using the Alhambra Faint Object Spectrograph and Camera (ALFOSC) on the 2.6~m Nordic Optical Telescope (NOT; \citep{2010ASSP...14..211D}) at 29~days and 50~days after discovery, the SPectrograph for the Rapid Acquisition of Transients \citep[SPRAT;][]{2014SPIE.9147E..8HP} on the 2~m Liverpool Telescope \citep[LT;][]{2004SPIE.5489..679S} at 43~days, the ESO Faint Object Spectrograph and Camera version 2 \citep[EFOSC2;][]{1984Msngr..38....9B} on the New Technology Telescope (NTT) at 131~days, 176~days and 249~days as part of the extended Public ESO Spectroscopic Survey of Transient Objects (ePESSTO; \citealt{2015A&A...579A..40S}), and the Optical System for Imaging and low-Intermediate-Resolution Integrated Spectroscopy+ \citep[OSIRIS+;][]{2000SPIE.4008..623C} on the 10.4~m Gran Telescopio Canarias (GTC) at 206~days. 

LCO optical spectra were taken with the FLOYDS spectrographs on the 2~m Faulkes Telescopes North and South (FTN/FTS), using a 2\arcsec\ slit. The raw spectra cover wavelengths 3500 to 10,000 \AA at a resolution $R \approx 300$--$600$. The spectra were reduced using the \texttt{floydsspec} pipeline\footnote{\url{https://github.com/svalenti/FLOYDS_pipeline/}}, which calibrates wavelength and flux, removes cosmic rays, and extracts the final spectrum, as detailed in \cite{Valenti2014}.
A log of optical spectroscopy of 2024aepd is presented in Table~\ref{tabc1}.

Dark Energy Spectroscopic Instrument (DESI) is amounted on the 4-meter Mayall Telescope, located at the Kitt Peak National Observatory in Arizona, USA \citep{DESI2016a,DESI2022b}. 
On February 12, 2024, DESI obtained one spectrum of 2024aepd. This spectrum was selected from the DESI Bright Galaxy Survey \citep{Hahn2023}, with target sources coming from the DESI Legacy Imaging Survey \citep{Dey2019,Zou2017}. The spectral range spans from 3600~\AA~ to 9800~\AA, with an exposure time of $496$~s and a spectral resolution of $R$~=~2000--5000. The observed data were processed using the DESI spectroscopic data processing pipeline \citep{Guy2023}.

\subsection{X-ray Observations}\label{sec:x_combination}
The X-ray observations with the X-Ray Telescope \citep[XRT;][]{2005SSRv..120..165B} on the Neil Gehrels Swift Observatory were obtained in photon-counting mode. We processed the XRT data using the online tools provided by the UK Swift Science Data Centre \citep{2007A&A...469..379E,2009MNRAS.397.1177E} to generate both light curves and spectra.
A total of 35 observations are included in this work. To improve the signal-to-noise ratio, we divided the XRT observations into five epochs, using a binning scheme that ensured at least 30 counts per epoch while keeping the counts comparable across epoch.
Each stacked spectrum was grouped to contain a minimum of three counts per bin, and the fits were performed in XSPEC using the Cash statistic \citep{1979ApJ...228..939C}.
We adopted a Galactic neutral hydrogen column density toward 2024aepd of $3.45\times10^{20}~\mathrm{cm}^{-2}$ from the HI4PI survey \citep{2016A&A...594A.116H}.

\subsection{Near-infrared Photometry}
We utilized photometric data from the Nordic Optical Telescope near-infrared Camera and spectrograph (NOTCam), Espectr$\rm \acute{o}$grafo Multiobjeto InfraRojo (EMIR; \citealt{Garzon2022}) infrared camera/spectrograph at the GTC, and the Sun Yat-sen University (SYSU) 80\,cm infrared telescope \citep{Dong2026} to investigate the NIR emission of 2024aepd.

We obtained seven epochs of NIR imaging in the $JK$ bands with the NOTCam instrument mounted on the NOT. The NOTCam data were reduced using a slightly modified version of the NOTCam quicklook v2.5 reduction package. The reduction process included flat-field correction, a distortion correction, bad pixel masking, sky subtraction and finally stacking of the dithered images. PSF photometry on the NIR data was performed with the {\sc autophot} pipeline \citep{2022A&A...667A..62B} after template subtraction and the resulting magnitudes were calibrated using the 2MASS catalogue \citep{2006AJ....131.1163S}. For further details, see 
\citet{2026A&A...708A.139R}.  

For the $J$-band photometry, we made use of the high quality imaging from 2025-06-16 as a template. As this image is only $\sim150$ observer frame days after the first IR data, we tested using the latest available image from 2025-09-06 as a template as well, and found consistent results. We are therefore confident that any persistent $J$-band flux in the templates is too faint to be detectable in our templates. For the $K$-band photometry, we again made use of the high quality imaging from 2025-06-16 as a template, and found consistent results using 
the data from 2025-09-06. However, we can not absolutely rule out a late-rising echo that would contaminate our template. This would cause us to underestimate the true $K$-band fluxes. The TDE was detected only in the first three and two epochs for $J$ and $K$ respectively. 

GTC observations were obtained with EMIR in the $J$, $H$, and $K_s$ bands on 2025 October 1, $\sim300$~days after the optical discovery. In each filter, the target was observed with two 7-point dither sequences, with 5~s exposures at each dither position, yielding a total exposure time of 70~s per band. Two $K_s$-band frames affected by satellite trails were excluded before stacking. The raw frames were reduced with \texttt{PyEmir} \citep{Pascual2010,Cardiel2019}. Aperture photometry was performed with \texttt{Photutils} \citep{Photutils_19636730} after subtracting a two-dimensional background model and refining the source position through a local Gaussian centroid fit. Fluxes were measured within a circular aperture of radius $2\arcsec$, and photometric zero points were calibrated using isolated 2MASS field stars \citep{2MASS}.

Additional $J$-band monitoring was carried out with the SYSU 80~cm infrared telescope between 2025 September 5 and October 10, covering five epochs. Standard differential photometry shows that the total $J$-band emission remained approximately constant over this $\sim35$-day interval, with peak-to-peak variations of no more than $\sim$0.2~mag.

We also analyzed archival images obtained with the Wide Field Camera (WFCAM) on the United Kingdom Infrared Telescope (UKIRT) as part of the UKIRT Hemisphere Survey (UHS), retrieved from the WFCAM Science Archive \citep{WFCAM,UKIRT_UHS}. The available data comprise $J$-, $H$-, and $K$-band observations acquired on 2013 April 22, 2022 March 7, and 2019 July 16, respectively. Aperture photometry was performed with \texttt{Photutils} using a fixed $2\arcsec$-radius aperture centered on the source position, with photometric zero points taken from the image headers.

\subsection{Radio Observations}
We observed 2024aepd with enhanced Multi-Element Radio Linked Interferometer Network (e-MERLIN) in $C$ band on 2025 September 11 (MJD 60929.61; program RR20002), covering 4.82--5.33\,GHz. The data were reduced with the e-MERLIN CASA Pipeline (v1.1.19) in Common Astronomy Software Applications package (CASA v5.8.0; \citealt{CASA}). We used 3C~286, OQ~208, 1524+1521, and 3C~84 as the flux density, bandpass, phase, and point-source calibrators, respectively. Imaging was performed with \texttt{tclean} using Briggs weighting (\texttt{robust}=0.5). No significant radio emission was detected at the transient position, yielding a $3\sigma$ upper limit flux of 0.09~mJy, corresponding to a $3\sigma$ upper limit luminosity of $\sim$$8.42\times10^{37}~\mathrm{erg~s^{-1}}$.

We also observed 2024aepd with the Australia Telescope Compact Array (ATCA) on 2026 March 15 (MJD 61114.74) under program C3615 (PI: Yanan Wang). The observation was conducted in the 6D configuration using four 1.92-GHz-wide continuum bands centred at 5.1, 7.2, 9.2, and 11.3~GHz. The data were calibrated in CASA 6.7.2, with 1934$-$638 used to set the flux-density scale and calibrate the bandpass and 1502+106 used to calibrate the complex gains. The two lower-frequency bands and the two higher-frequency bands were combined separately to produce images at effective frequencies of 6.25 and 10.29\,GHz, respectively. Imaging was performed using multi-term multi-frequency synthesis deconvolution with \texttt{nterms}=2 \citep{mtmfs} and Briggs weighting with \texttt{robust}=1.0. No significant emission was detected at the position of 2024aepd, yielding $3\sigma$ upper limit fluxes of 28.7~$\mathrm{\mu Jy}$ and 27.1~$\mathrm{\mu Jy}$,
with the $3\sigma$ upper limit luminosities of $\sim$$3.36\times10^{37}~\mathrm{erg~s^{-1}}$ and $\sim5.22\times10^{37}~\mathrm{erg~s^{-1}}$ at 6.25 and 10.29~GHz, respectively.

\section{Results}
\label{sect:results}

\subsection{Black hole mass measurement and host-galaxy properties}
We analysed the DESI spectrum obtained 297~days prior to the TDE discovery. Following the spectral fitting procedure described in \citet{2023ChPhB..32c9801L}, we first matched the spectral resolution of the DESI data to that of the \citet{Bruzual_Charlot_2003} simple stellar population templates by convolving the observed spectrum with a Gaussian kernel, thereby degrading the higher DESI resolution to the template resolution. The stellar continuum was then modelled with a linear combination of model spectra, and the line-of-sight velocity distribution was characterized by an additional Gaussian convolution $G(\sigma_*)$. Strong emission lines were iteratively masked during the fit to avoid contamination of the continuum. This yields a velocity dispersion of $79.6\pm6.5~\mathrm{km\,s^{-1}}$.
Using the $M_{\rm BH}-\sigma_*$ relation of \citet{2013ARA&A..51..511K}, our velocity dispersion measurement yields a black hole mass of $\log_{10}(M_{\rm BH}/M_{\odot})=6.74\pm0.33$, where the uncertainty includes the intrinsic scatter in the $M_{\rm BH}-\sigma_*$ relation. Alternatively, the black hole mass was also estimated using the Modular Open Source Fitter for Transients \citep[MOSFiT;][]{2018ApJS..236....6G,2019ApJ...872..151M}, with photometry from the ATLAS $c$ and $o$ bands. Adopting the nested sampling method yielded $\log_{10}(M_{\rm BH}/M_{\odot})=6.30^{+0.20}_{-0.26}$, consistent with the prior estimate, and $M_{*}/M_{\odot}=0.33^{+0.18}_{-0.12}$. The detailed fitting outputs are provided in Table~\ref{tabA1}, Figs.~\ref{figA1} and \ref{figA2}. 

We modeled the host SED with the \texttt{Prospector} \citep{Prospector}, using the flexible stellar population synthesis module (FSPS; \citealt{2009ApJ...699..486C,2010ApJ...712..833C}). 
The optical-to-IR photometry was adopted from public survey catalogs. Specifically, we used the SDSS DR~18 \texttt{modelMag} measurements in $u$, $g$, $r$, $i$, and $z$ \citep{SDSS18}, the 2MASS All-Sky Point Source Catalog default magnitudes in $J$, $H$, and $K_s$ \citep{2MASS}, and the AllWISE Source Catalog profile-fit magnitudes (\texttt{w1mpro}, \texttt{w2mpro}, and \texttt{w3mpro}) in $W1$, $W2$, and $W3$ \citep{WISE}. 
For the UV bands, we performed aperture photometry on archival \textit{GALEX} far-UV (FUV) and near-UV (NUV) data with \texttt{gPhoton} (\citealt{GALEX}; \citealt{gphoton}), using an $8^{\prime\prime}$ radius aperture. Moreover, we assumed a delayed exponential star formation history, a initial mass function of \citet{Chabrier2003}, and a attenuation law of \citet{Calzetti2000}. We sampled the posterior
distribution with the \texttt{dynesty} nested sampler \citep{2020MNRAS.493.3132S,koposov_dynesty_zenodo}. Overall, the fitting yields a reduced $\chi^2$ of 0.57. The best-fitting SED is shown in Fig.~\ref{fig:host_sed}. 

The inferred stellar mass is $10^{9.81_{-0.06}^{+0.05}}\,M_\odot$, and the instantaneous star-formation rate is $0.09_{-0.08}^{+0.05}\,M_{\odot}\,{\rm yr}^{-1}$, consistent with no ongoing star formation.
These parameters indicate that the host is a moderately massive, quiescent galaxy, in agreement with its pre-TDE absorption-dominated spectrum. It is worth noting that a contaminating source is located only $3.7"$ from 2024aepd. Therefore, the absolute values of these quantities should be taken with caution.

\begin{figure}
    \centering
    \includegraphics[width=0.7\textwidth]{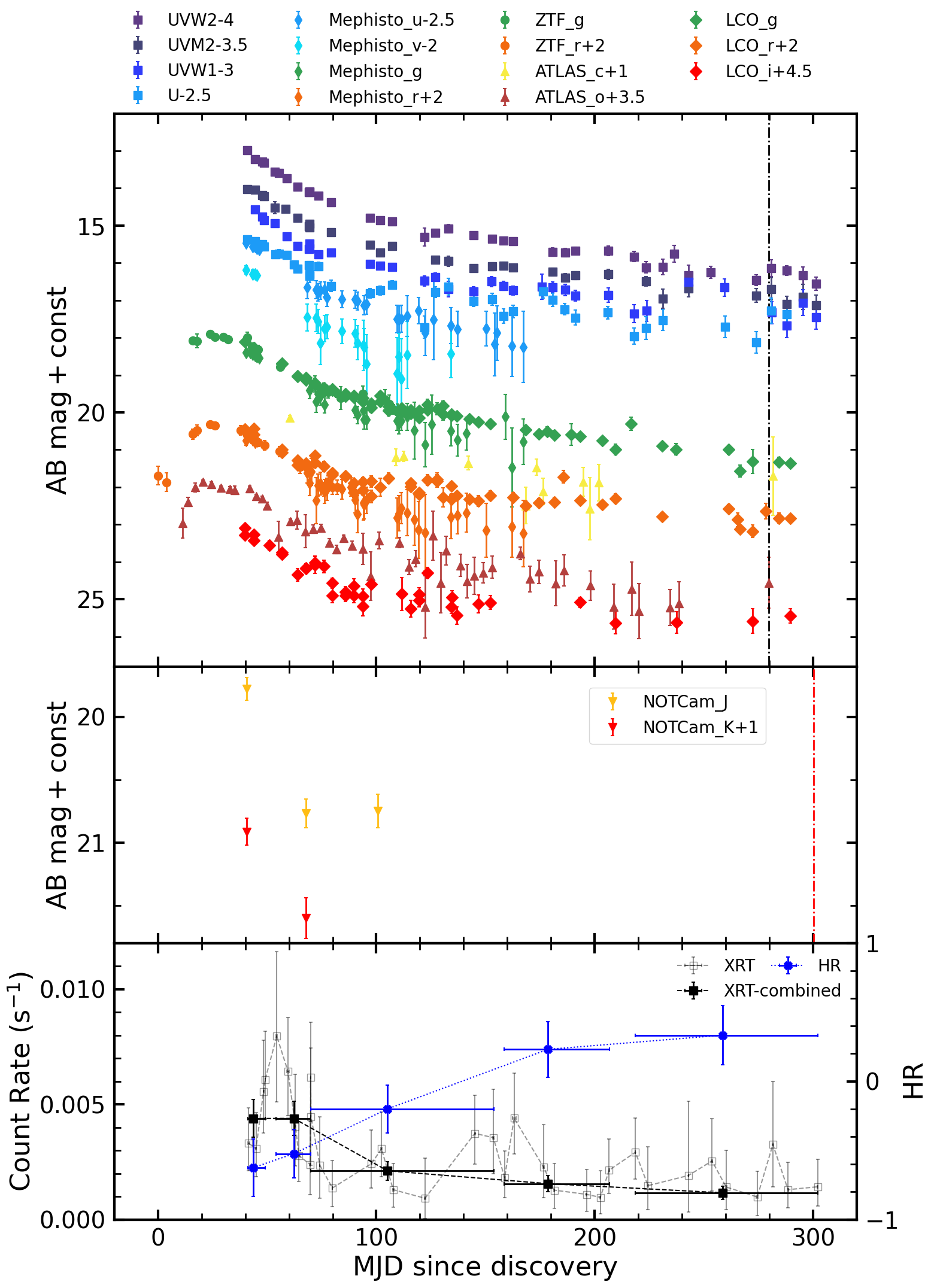}
    \caption{Multi-wavelength light curves of 2024aepd. 
    Top: Host-subtracted and galactic-extinction-corrected UV--optical light curves obtained with UVOT, Mephisto, ZTF, ATLAS and LCO. The black vertical dash-dotted marks the epoch of the e-MERLIN observation. Middle: NOTCam $J$- and $K$-band light curves. The red vertical dash-dotted line marks the epoch of the GTC observation. Lower: X-ray light curve and the hardness ratio measured by XRT. The gray open squares show count rates from individual observations, while the black squares indicate count rates derived from the stacked observations. }
    \label{fig1}
\end{figure} 

\begin{figure}
    \centering
    \includegraphics[width=0.7\textwidth]{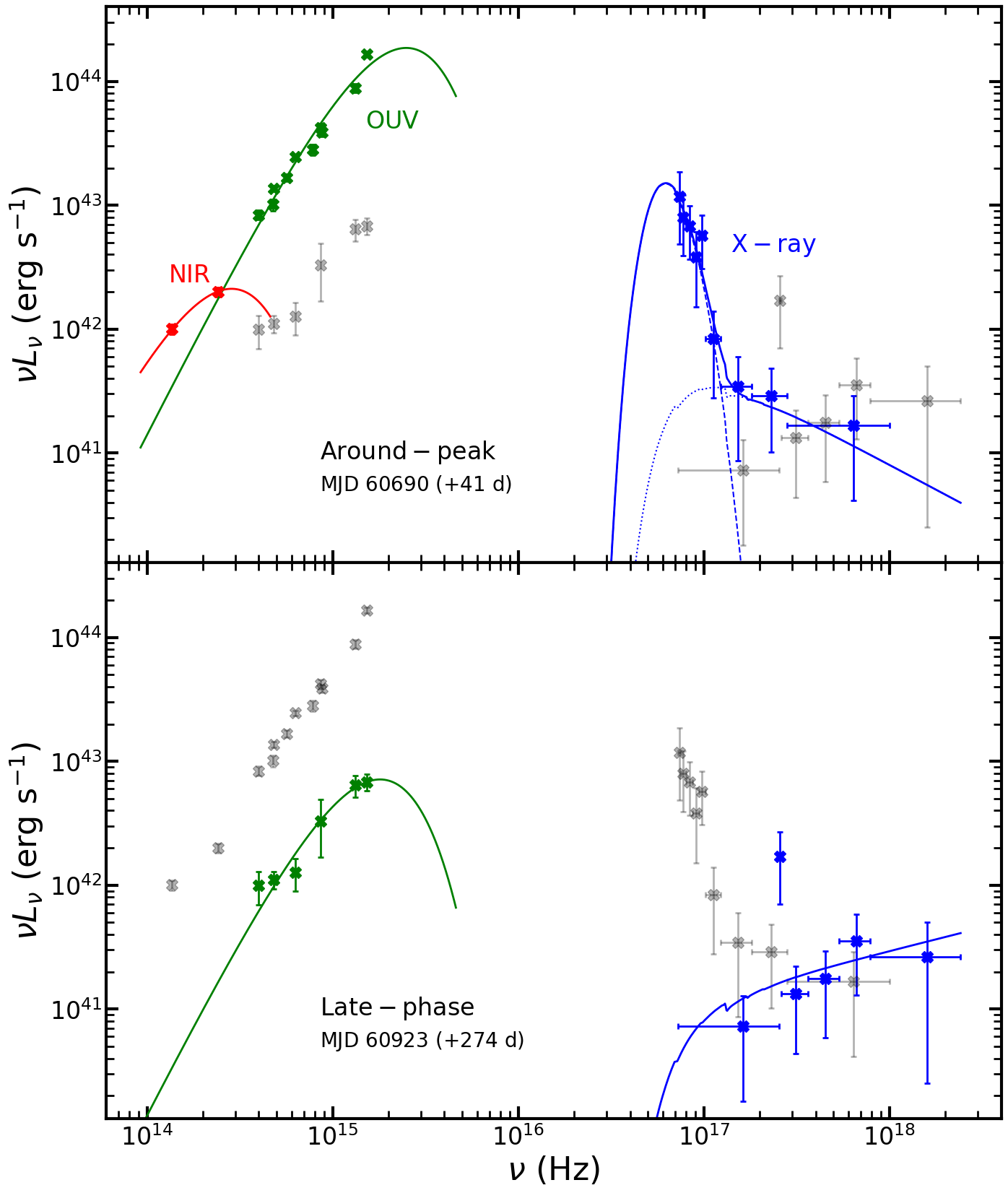}
    \caption{The NIR-to-X-ray SEDs constructed from nearly simultaneous observations obtained near peak and during the late post-peak phase. The light red, green, and blue crosses represent the NIR, optical/UV, and X-ray data, respectively. Data from different epochs are shown in grey for comparison. The red and green solid lines represent the best-fitting single-temperature blackbody models for the NIR and optical/UV emission, respectively, while the blue solid lines represent the best-fitting models for the X-ray emission. In the upper panel, the blue dashed and dotted lines represent the \texttt{diskbb} and \texttt{powerlaw} components of the X-ray emission, respectively.}
    \label{fig2}
\end{figure} 
\subsection{Multi-band photometry}
Fig.~\ref{fig1} presents the photometric evolution of 2024aepd. The ZTF and ATLAS light curves exhibit an initial rise, reaching a common peak around MJD~60670--60673. After maximum light, the UV-to-optical emission declines steadily, with the bluer bands fading faster than the redder ones. The NIR observations, however, only sample the declining phase of the light curve, preventing any direct constraint on the relative timing of the optical and NIR peaks.

In the X-ray band, short-term variability suggestive of flaring activity is superimposed on the overall long-term decline. In particular, a prominent flare is detected near the beginning of our X-ray monitoring campaign, starting at $\sim$45~days after discovery and lasting for approximately 50~days.
Given the subsequent low signal-to-noise ratio of individual observations, we group the data into five epochs (see details in Sec.~\ref{sec:x_combination}) and compute the hardness ratio for each combined epoch, defined as $(H-S)/(H+S)$, where $H$ and $S$ are the photon counts in the 1.0--10~keV and 0.3--1.0~keV bands, respectively. The hardness ratio increases gradually over the course of the Swift monitoring, indicating progressive spectral hardening in 2024aepd.

\subsection{Broadband spectral energy distribution}
In Fig.~\ref{fig2}, we present two representative broadband SEDs spanning the NIR to X-ray bands. The top panel shows the early-phase SED near peak brightness at MJD~60690 (+41 days). It combines the stacked XRT spectrum in the X-ray band, the NOTCam observations in the NIR, and contemporaneous UV--optical photometry from UVOT, Mephisto, and LCO.
For comparison, the bottom panel shows the late-phase SED at MJD~60873 (+224 days). Because no contemporaneous Mephisto observations are available at this epoch, we use the UVOT and LCO data to characterize the UV--optical emission, while the NIR counterpart is no longer detected. As the TDE evolves, the broadband luminosity declines substantially across the NIR, UV--optical, and X-ray bands. At all epochs, the observed SED requires multiple emission components, with at least one component contributing to each spectral regime.

\begin{figure}
    \centering
    \includegraphics[width=0.7\textwidth]{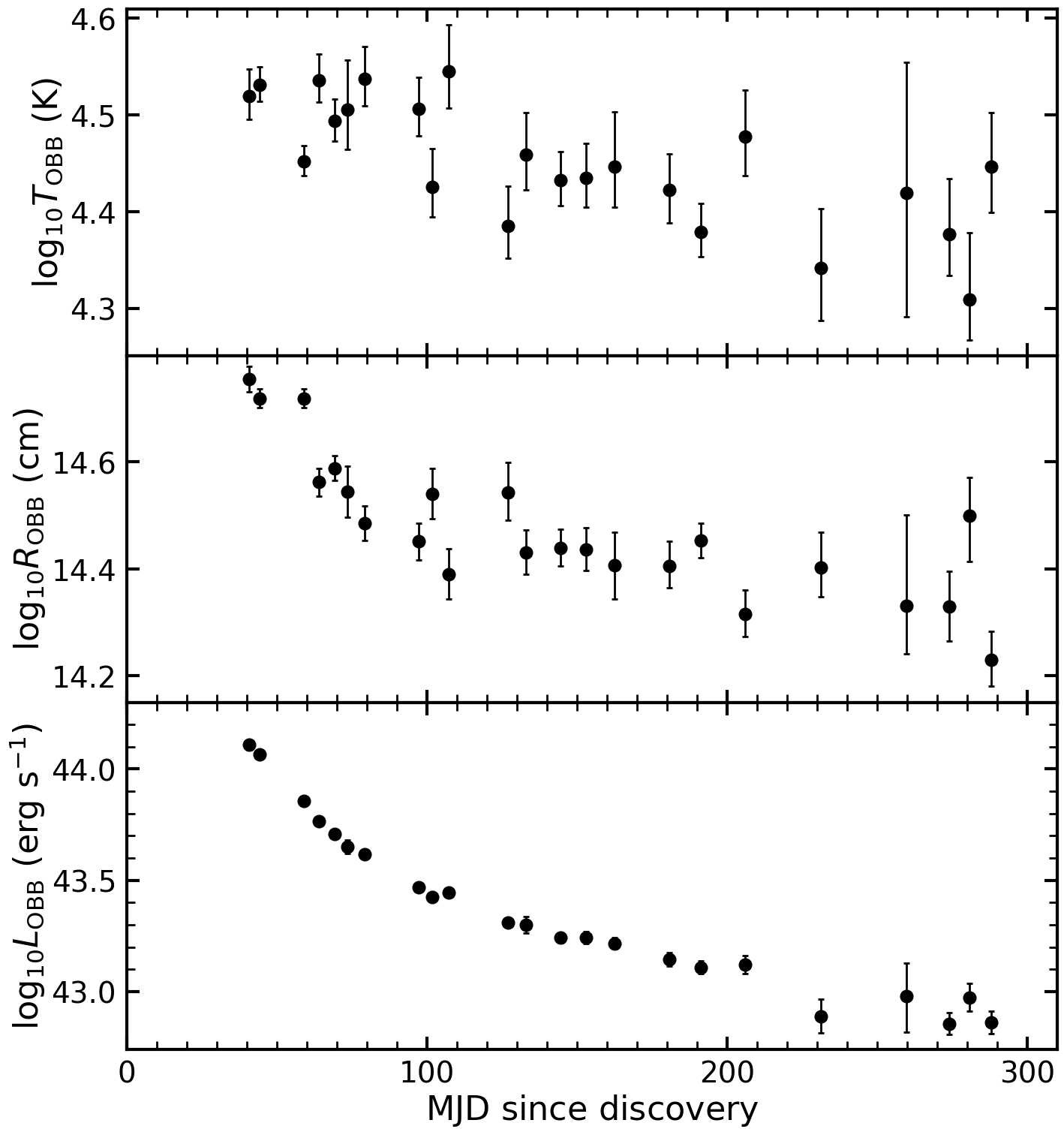}
    \caption{The temporal evolution of the blackbody temperature (top panel), radius (middle panel) and luminosity (bottom panel) of the UV--optical emission.}
    \label{fig3}
\end{figure} 

\begin{figure}
    \centering
    \includegraphics[width=0.7\textwidth]{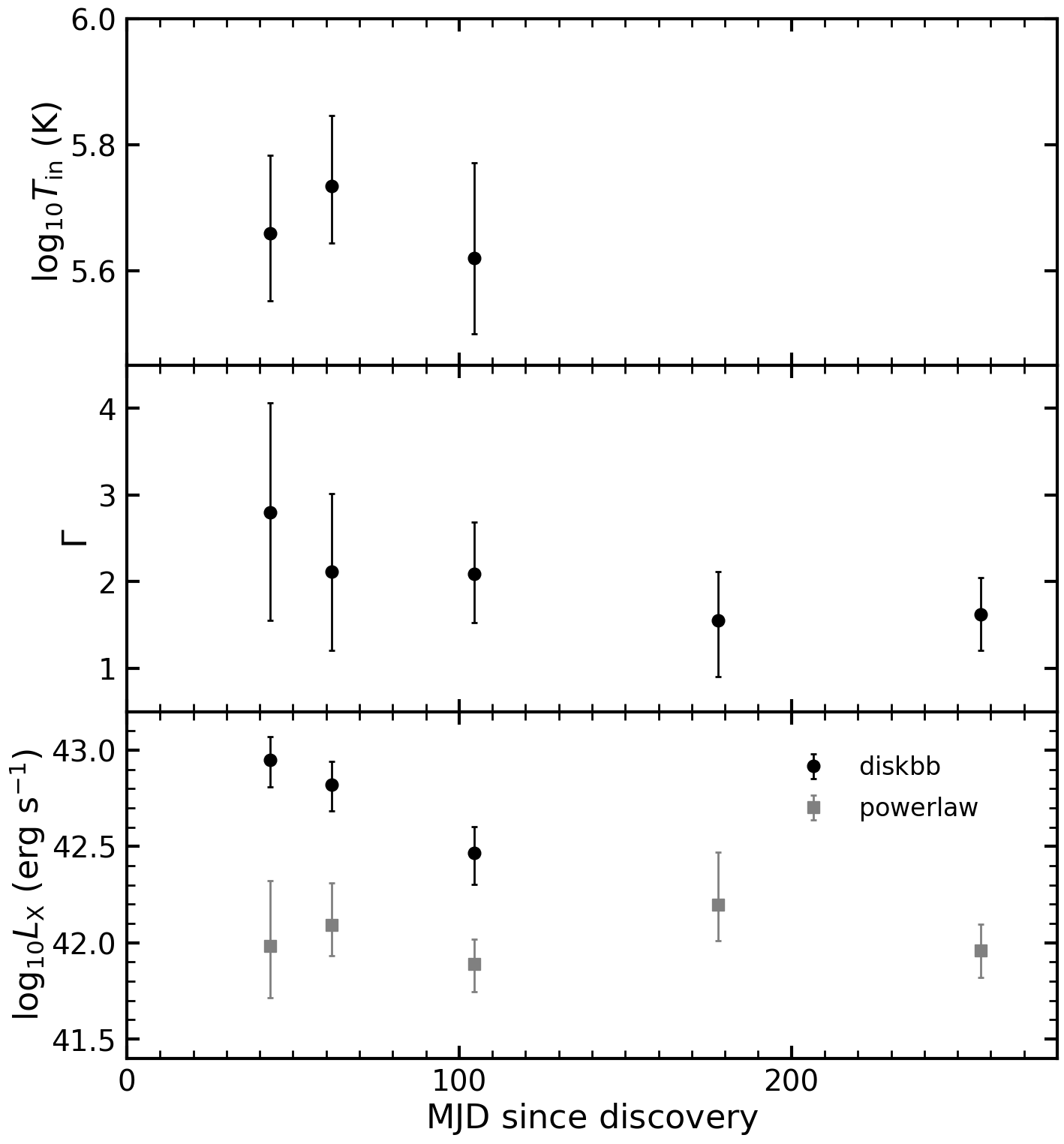}
    \caption{The temporal evolution of the disk inner temperature $T_\mathrm{in}$ of \texttt{diskbb} component (top panel), the photon index $\Gamma$ of the \texttt{powerlaw} component (middle panel), and the 0.3--10~keV luminosities of the \texttt{diskbb} and \texttt{powerlaw} components.}
    \label{fig4}
\end{figure}
To construct the UV--optical SED, we select, for each UVOT epoch, the temporally closest Mephisto and LCO observations, requiring a separation of less than three days. Three epochs are excluded, since the $i$-band photometry at $\sim122$ days, $r$-band photometry at $\sim186$ days, and $g$-band photometry at $\sim218$ days of the LCO deviate significantly from the overall declining trend.
We model the UV--optical SED with a single blackbody component using a Markov Chain Monte Carlo (MCMC) approach, as shown by the green line in Fig.~\ref{fig2}. Fig.~\ref{fig3} presents the temporal evolution of the best-fitting blackbody temperature $T_\mathrm{OBB}$, radius $R_\mathrm{OBB}$, and luminosity $L_\mathrm{OBB}$.
$L_\mathrm{OBB}$ peaks at $\sim 1.3\times10^{44}$~erg~s$^{-1}$ and subsequently declines as a power-law, $L_\mathrm{OBB}\propto t^{-1.52\pm0.04}$, throughout the studied period. The temperature decreases significantly from $\sim3.3\times10^{4}$ to $\sim2.4\times10^{4}~\mathrm{K}$, following $T_{\rm OBB}\propto t^{-0.16^{+0.03}_{-0.02}}$. The blackbody radius reaches a maximum of $\sim6\times10^{14}~\mathrm{cm}$ and subsequently declines as $R_{\rm OBB}\propto t^{-0.55\pm0.03}$.

\begin{figure}
    \centering
    \includegraphics[width=\columnwidth]{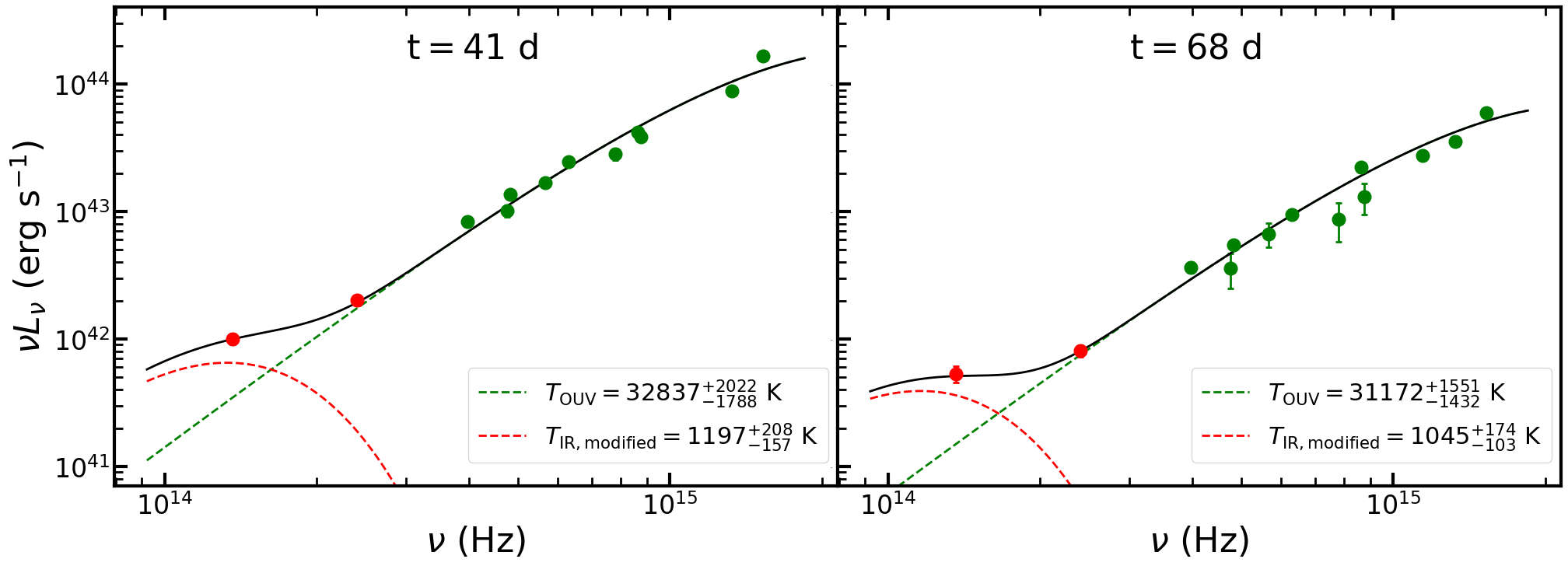}
    \caption{Best-fitting results for the two UV--optical--NIR SED at 41~days and 68~days using modified dust blackbody with an additional the UV--optical blackbody. The red and green points represent the NOTCam and the UV--optical observations, respectively. The red dotted lines represent the modified blackbody for the NIR excess emission and the green dotted lines represent the UV--optical blackbody. The combined models are plotted as the black solid lines. We also plotted the best-fitting temperatures for the two components with their uncertainties.}
    \label{fig5}
\end{figure}

At early times, both the NOTCam $J$- and $K$-band measurements show excess emission above the Rayleigh--Jeans tail of the UV--optical blackbody, as illustrated in the upper panel of Fig.~\ref{fig2}. A commonly invoked explanation for the IR excess observed in TDEs is reprocessed emission from pre-existing circumnuclear dust, often referred to as a dust echo \citep{2016MNRAS.458..575L,2016ApJ...829...19V,2016ApJ...828L..14J}. 
To investigate the IR excess of 2024aepd, we construct the UV--optical--NIR SEDs with the first two NOTCam epochs and quasi-simultaneous UV/optical observations. We then fit the SEDs using a single-temperature blackbody for UV--optical emission and a modified blackbody, $f_\nu\propto Q_\nu B_\nu(T)$, for the emission of the optically thin dust, where $Q_\nu$ is the absorption efficiency. Following \citet{2016ApJ...829...19V}, we adopt $Q_\nu \propto \nu^{1.8}$, corresponding to the characteristic emissivity law of 0.1 $\mu $m graphite grains. The best-fitting models for the two epochs are shown in Fig.~\ref{fig5}. The inferred IR temperature remains consistent at $\sim1100$~K between 41 and 68~days.
We also fit the NIR SEDs with a power-law model, $f_\nu\propto\nu^{\alpha_{\rm IR}}$. The resulting spectral indices, $\alpha_{\rm IR}=0.22^{+0.23}_{-0.22}$ and $-0.27^{+0.33}_{-0.31}$, are both consistent with zero, indicating an approximately flat NIR spectrum. 

In the X-ray band, we first fit the spectrum with an absorbed multi-temperature blackbody model, i.e., \texttt{tbabs*zashift*diskbb}. We adopt the \texttt{tbabs} component to account for the interstellar absorption, using the \textit{abund wilm} command to set the abundance table \citep{2000ApJ...542..914W} and \textit{xsect vern} command to set the photoelectric cross sections \citep{1996ApJ...465..487V}. This yields a clear hard tail in the residuals, indicating the presence of an additional high-energy component. We find no evidence for intrinsic absorption in excess of the Galactic column density. Therefore, we fix the hydrogen column density at the Galactic value of $3.45\times10^{20}~\mathrm{cm}^{-2}$ \citep{2016A&A...594A.116H}.
We then add a \texttt{powerlaw} component to the model, which significantly improves the fit during the first three epochs. In contrast, for the remaining epochs, an absorbed \texttt{powerlaw} model alone provides an adequate description of the spectrum, with no statistically significant improvement obtained by including an additional \texttt{diskbb} component, as assessed using the {\sc ftest} command.
The best-fitting results are shown in Fig.~\ref{fig4}. 
As the \texttt{diskbb} luminosity decreases from $8.91^{+2.42}_{-2.89}\times10^{42}~\mathrm{erg~s^{-1}}$ to $2.93^{+0.91}_{-1.12}\times10^{42}~\mathrm{erg~s^{-1}}$, the \texttt{diskbb} temperature $T_\mathrm{in}$ varies within errors. At the same time, $\Gamma$ exhibits a tentative decreasing trend, while the \texttt{powerlaw} luminosity remains nearly constant.

To assess whether our results depend on the choice of thermal and non-thermal spectral components, we combine all exposures into a single spectrum and fit it with several model combinations. Specifically, we consider two thermal models, \texttt{diskbb} and \texttt{tdediscspec} \citep{2021MNRAS.507L..24M}, and two non-thermal models, \texttt{powerlaw} and \texttt{simpl} \citep{2009PASP..121.1279S}. The best-fitting parameters are listed in Table~\ref{tabB1}. We find that the inferred spectral parameters are largely insensitive to the adopted model combination. The only notable difference lies in the characteristic disk temperature: the \texttt{diskbb} fit yields $T_{\rm in}\sim1.8T_{\rm p}$, where $T_{\rm p}$ denotes the maximum disk temperature defined in \texttt{tdediscspec}.

\subsection{Spectroscopic Evolution}
\begin{figure}
    \centering
    \includegraphics[width=0.7\textwidth]{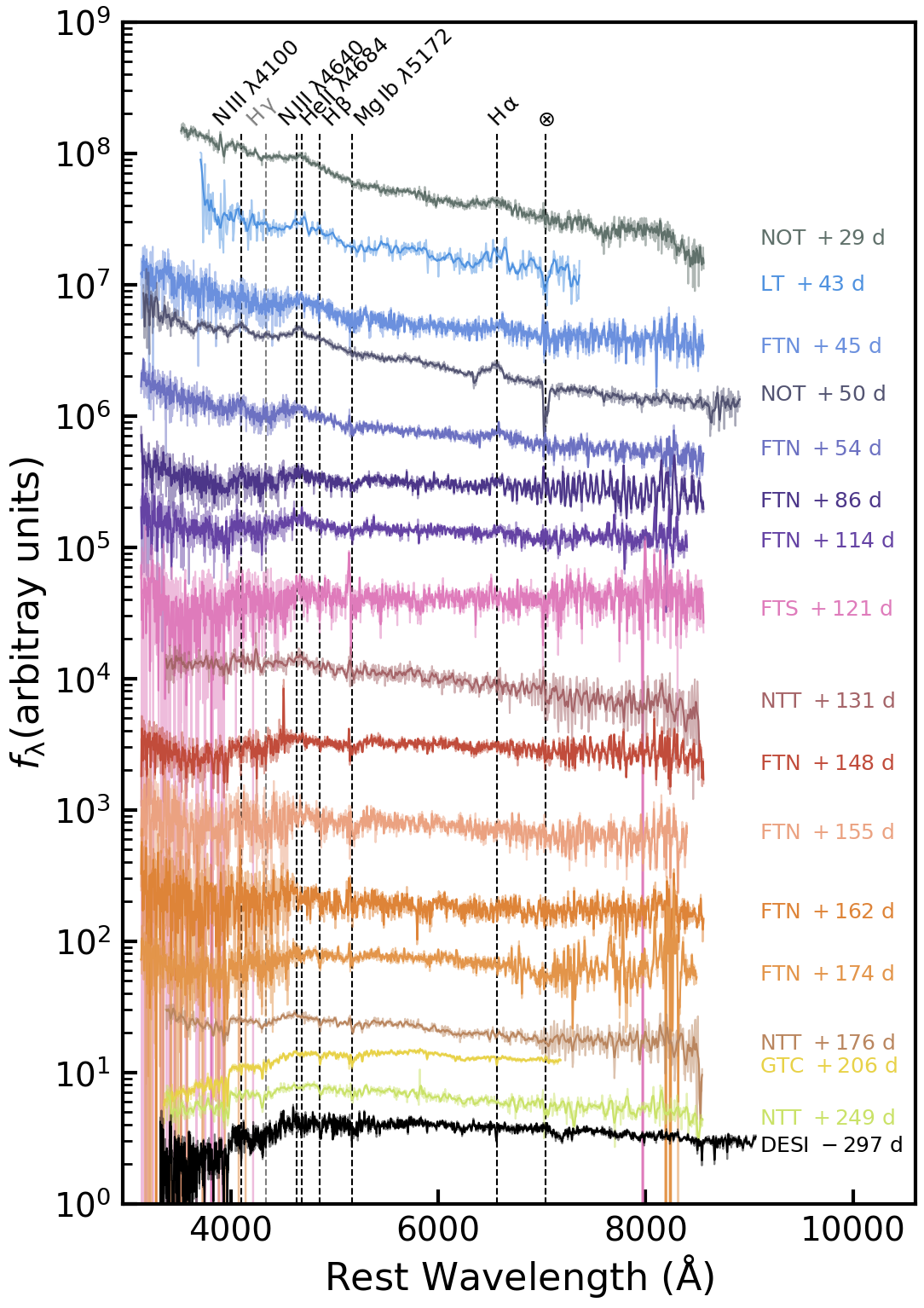}
    \caption{Optical spectroscopic evolution of 2024aepd. The vertical lines mark prominent observed emission and absorption spectral features. The bottom black spectrum represent the spectrum of the host galaxy taken about 300~days before TDE discovery. $\otimes$ marks telluric absorption features.}
    \label{fig6}
\end{figure}

Fig.~\ref{fig6} shows the spectral evolution of 2024aepd from $\sim$30 to $\sim$250 days after discovery, with the most prominent spectral features labeled. The bottom spectrum represents the host galaxy and was obtained approximately 300 days before the TDE discovery. It shows H\,$\alpha$, H\,$\beta$, and Mg\,Ib triplet absorption features, with no evidence of emission lines.

The first NOT spectrum, obtained 29~days after discovery, exhibits a blue continuum together with broad H$\alpha$ emission and a blend of He\,II $\lambda4686$ and N\,III $\lambda4640$. In the subsequent two spectra, both the Balmer and He\,II emission features become more prominent relative to the continuum, classifying 2024aepd as an ``H+He'' TDE. The emission feature near 4100\,\AA\ could be associated with either H$\delta$ or N\,III. However, given the absence of H$\gamma$ emission, we tentatively identify this feature as N\,III.

By day 131, both the blue continuum and the emission lines have weakened significantly. In the subsequent spectra, the He\,II emission continues to fade, while the Balmer features become increasingly dominated by absorption. Although the final two spectra are largely dominated by host-galaxy absorption features, the blue continuum remains noticeably brighter than that of the host galaxy, indicating that the TDE emission is still present at these late epochs.

Moreover, after day 54, the Mg\,Ib absorption triplet becomes identifiable again. An additional emission feature is also present near 5150\,\text{\AA}, which may be associated with either [Fe\,VI] $\lambda5145$ or [Fe\,VII] $\lambda5159$. The absence of the accompanying [Fe\,VII] lines at $\lambda5721$ and $\lambda6087$ may favor the [Fe\,VI] identification. However, a more detailed analysis is beyond the scope of the present work.

\section{Discussion}
\label{sect:discussion}

In this work, we present multi-wavelength observations of the 2024aepd from the IR to soft X-rays. The event exhibits several notable features, including a prominent NIR excess with an approximately flat spectrum, whose slope remains consistent within the uncertainties between the two epochs at 41 and 68~days after discovery. We also detect the early emergence of a hard X-ray tail superimposed on a dominant thermal component, followed by a transition from a thermal-dominated to a non-thermal-dominated X-ray spectrum at $\sim178$ days. During this interval, the X-ray spectrum gradually hardens, with the hardness ratio increasing from $\sim-0.6$ to $\sim0.3$, while the luminosity declines by approximately one order of magnitude. We discuss the implications of these results and compare them with other TDEs below.

\subsection{Origin of IR emission}
\begin{figure}
    \centering
    \includegraphics[width=0.7\columnwidth]{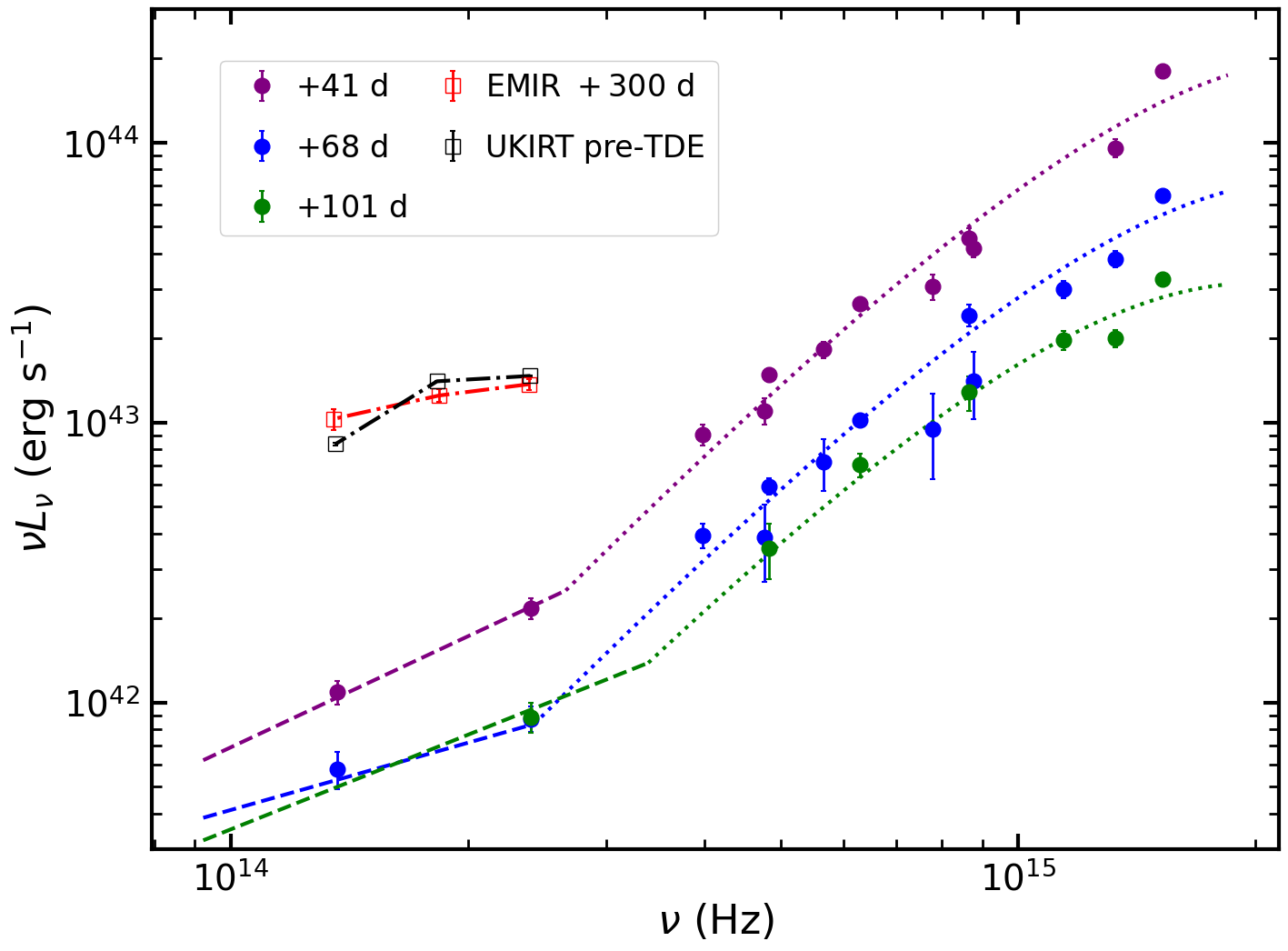}
    \caption{Best-fitting models for the UV--optical--NIR SEDs at different phases. The dotted and dashed lines represent the truncated UV--optical blackbody and the free--free emission component at longer wavelengths, respectively. The green lines show the fit to the SED at 101~day, supplemented with the $K$-band measurement at 68~day. For comparison, the EMIR observations obtained at $\sim300$~days, which include the host-galaxy contribution, and the pre-TDE UKIRT observations are shown as red and black open squares connected by dash-dotted lines, respectively.}
    \label{fig7}
\end{figure}

In Fig.~\ref{fig7}, we present the host-subtracted NOTCam observations together with the quasi-simultaneous UV--optical data. To assess the late-time NIR emission, we also include the unsubtracted EMIR observations obtained at $\sim300$~days and the pre-TDE UKIRT measurements. The late-time EMIR $J$- and $H$-band fluxes are slightly lower than their corresponding pre-TDE UKIRT values, whereas the EMIR $K_s$-band flux is significantly higher. Given these opposite trends, the apparent differences may be dominated, at least in part, by systematic uncertainties in the photometry. 

To investigate the origin of the NIR emission, we first consider the conventional dust-echo scenario.
Following \citet{2016ApJ...829...19V}, we calculate the sublimation radius using
\begin{equation}
\begin{aligned}
    R_\mathrm{sub}=0.15\left(\frac{L_\mathrm{45}}{a_\mathrm{0.1}^2T_\mathrm{1850}^{5.8}}\right)^{1/2}~\mathrm{pc},
\end{aligned}
\end{equation}
where $L_\mathrm{45}$ is the bolometric luminosity in units of $10^{45}~\mathrm{erg~s^{-1}}$, $a_\mathrm{0.1}$ is the grain size in units of 0.1~$\mu$m, and $T_\mathrm{1850}$ is the temperature of the dust in units of 1850~K. We estimate the unabsorbed peak pseudo-bolometric luminosity by summing the UV--optical blackbody luminosity and the luminosities of the \texttt{diskbb} and \texttt{powerlaw} components. Based on the ratio shown in Fig.~\ref{fig4}, we assume that the \texttt{powerlaw} luminosity is equal to 10\% of the \texttt{diskbb} luminosity. This yields a peak pseudo-bolometric luminosity of $1.38\times10^{45}~\mathrm{erg~s^{-1}}$, substantially higher than the measured $L_{\rm OBB}$ and $L_{\rm X}$. This difference arises from the relatively low disk temperature of 40~eV, which places most of the disk emission in the unobservable EUV band. Assuming no missing emission and adopting a dust grain size of $a=0.1~\mu$m, we infer a sublimation radius $R_\mathrm{sub}\sim0.62~\mathrm{pc}$, corresponding to to a light-travel time of approximately 743~days.  
Because the luminosity inferred from the observed bands likely underestimates the line-of-sight bolometric luminosity, the corresponding sublimation radius, and hence the expected echo delay, should be even larger. By contrast, the characteristic dust radius can also be constrained from the light-travel time between the primary emitting region and the dust. Taking the first NIR observation as an approximate upper limit on the epoch of the IR peak yields $\tau<17$~days. 
Thus, even the minimum echo timescale inferred from the sublimation radius is substantially longer than the observational upper limit. This discrepancy is difficult to reconcile with a conventional dust-echo scenario. Nevertheless, anisotropic TDE emission, which may depend strongly on viewing angle, could cause the dust to receive a lower irradiating luminosity than implied by the isotropic-equivalent luminosity, thereby reducing the inferred sublimation radius and the expected echo delay. In addition, if the emitting dust lies close to the line of sight, the observed delay can be much shorter than the physical propagation time from the central source to the dust. In this geometry, the dust emission is produced approximately $R_{\rm sub}/c$ after the central UV--optical emission in the source frame, yet the shorter travel distance from the dust to the observer can cause the two signals to arrive with little or no measurable lag. We therefore cannot completely rule out a dust-echo origin.

Alternatively, motivated by the theoretical calculations of \citet{2020MNRAS.492..686L} and \citet{2020SSRv..216..114R}, we consider free--free emission from the reprocessing layer surrounding the accretion flow as another possible origin of the NIR excess. 
This interpretation has also been invoked to explain the NIR excesses observed in AT2019azh \citep{2026A&A...708A.139R} and TDE~2025abcr \citep{patra2026jwstkeckobservationsoffnuclear}. 

In an electron-scattering-dominated atmosphere ($\kappa_{\rm es} \gg \kappa_{\rm ff}$), the free--free opacity decreases strongly with increasing frequency. Consequently, photons of different frequencies emerge from different thermalization depths. While optical and UV photons are thermalized relatively deep within the reprocessing layer, lower-frequency NIR photons are thermalized at progressively larger radii because of their higher free--free opacity. The larger thermalization radius corresponds to a larger effective emitting area, boosting the NIR luminosity above the level expected from a single-temperature blackbody fitted to the optical/UV emission. Such frequency-dependent thermalization therefore provides a natural explanation for the observed NIR excess.

To quantify this effect, we follow \citet{2025ApJ...995..228S} and consider a homogeneous spherical medium in which electron scattering dominates the total opacity. We adopt $\kappa_{\rm es}=0.4\ {\rm cm^2\,g^{-1}}$ for pure hydrogen gas, while the free--free absorption coefficient $\kappa_{\rm ff}$ determines the thermalization depth. In the Rayleigh--Jeans limit, neglecting the Gaunt factor and taking $Z=1$, $\kappa_{\rm ff}$ is given by 
\begin{equation}
\begin{aligned}
    \kappa_\mathrm{ff}&=0.018T^{-3/2}\nu^{-2}n_e n_i \rho^{-1}\\
    &=0.018T^{-3/2}\nu^{-2}\rho m_p^{-2}\ [\mathrm{cm^{2}\,g^{-1}}],
\end{aligned}
\label{eq1}
\end{equation}
where $T$ and $\rho$ are the temperature and the density of the emitting region, $\nu$ is the frequency, $n_e$ and $n_i$ are the number density of the electrons and the ions, and $m_p$ is the proton mass.

We further assume a power-law density profile, $\rho=\rho_0(r/r_0)^{-s}$ with $r_0=10^{15}$~cm. Under the condition $\kappa_{\rm es}\gg \kappa_{\rm ff}$, the effective opacity is $\kappa_{\rm eff}=(\kappa_{\rm es}\kappa_{\rm ff})^{1/2}$, and the thermalization radius $r_{{\rm th},\nu}$ is defined by 
\begin{equation}
\begin{aligned}
    \tau_\mathrm{eff}=\kappa_\mathrm{eff}\rho r= (\kappa_\mathrm{es} \kappa_\mathrm{ff})^{1/2}\rho r_\mathrm{th,\nu}=1.
\end{aligned}
\label{eq2}
\end{equation}

Because electron scattering dominates the total opacity, the observed spectrum can be approximately described by the chromatic radiative diffusion solution \citep{1972Ap&SS..19...61I}, written in the form of  
\begin{equation}
\begin{aligned}
    \nu L_\nu=4\pi r_\mathrm{th,\nu}^2\frac{4\pi\nu B_\nu(T(r_\mathrm{th,\nu}))}{\kappa_\mathrm{es}\rho(r_\mathrm{th,\nu})r_\mathrm{th,\nu}(s-1)}.
\end{aligned}
\label{eq3}
\end{equation}
Substituting Eqs.~\ref{eq1} and \ref{eq2} into Eq.~\ref{eq3}, together with the density profile, yields 
\begin{equation}
\begin{aligned}
    \nu L_\nu=&\frac{32\pi^2 k_\mathrm{B}}{\kappa_\mathrm{es}c^2(s-1)}(\frac{0.018\kappa_\mathrm{es}}{m_\mathrm{p}^2})^{\frac{1+s}{3s-2}}\\
    &\times T^\frac{7-3s}{4-6s}(\rho_0r_0^s)^\frac{5}{3s-2}\nu^\frac{7s-8}{3s-2},
\end{aligned}
\label{eq4}
\end{equation}
where $k_\mathrm{B}$ is the Boltzmann constant.

According to this framework, there exists a characteristic frequency $\nu_\mathrm{break}$ such that photons with $\nu>\nu_\mathrm{break}$ can escape without significant absorption, corresponding to the UV--optical blackbody component. We represent this component using the best-fitting blackbody derived from the UV--optical SED, e.g.,
\begin{equation}
\begin{aligned}
    F_\nu=(\frac{R_\mathrm{OBB}}{D})^2\pi B_\nu(T_\mathrm{OBB}),
\end{aligned}
\label{eq5}
\end{equation}
where $D=394.4~\mathrm{Mpc}$ is the luminosity distance to the target.

In contrast, photons with $\nu<\nu_\mathrm{break}$ are absorbed and reprocessed along their propagation paths, producing the power-law component described by 
\begin{equation}
\begin{aligned}
    F_\nu=K(\rho_0,s,T)\nu^\frac{4s-6}{3s-2}/(4\pi D^2),
\end{aligned}
\label{eq6}
\end{equation}
where 
\begin{equation}
\begin{aligned}
    K(\rho_0,s,T)=&\frac{32\pi^2 k_\mathrm{B}}{\kappa_\mathrm{es}c^2(s-1)}(\frac{0.018\kappa_\mathrm{es}}{m_\mathrm{p}^2})^{\frac{1+s}{3s-2}}\\
    &\times T^\frac{7-3s}{4-6s}(\rho_0r_0^s)^\frac{5}{3s-2}.
\end{aligned}
\label{eq7}
\end{equation}
Requiring continuity of the flux at $\nu_\mathrm{break}$ leads to 
\begin{equation}
\begin{aligned}
    F_{\nu}&=(\nu/\nu_\mathrm{break})^\frac{4s-6}{3s-2} F_{\nu_\mathrm{break}}\\
    &=(\nu/\nu_\mathrm{break})^\frac{4s-6}{3s-2}(\frac{R_\mathrm{OBB}}{D})^2\pi B_{\nu_\mathrm{break}}(T_\mathrm{OBB}).
\end{aligned}
\label{eq8}
\end{equation}
The full UV--optical--NIR spectrum can then be described by 
\begin{equation}
    F_\nu=\left\{
    \begin{aligned}
        &(\frac{R_\mathrm{OBB}}{D})^2\pi\frac{2h\nu^3}{c^2} \frac{1}{e^{h\nu/k_\mathrm{B} T_\mathrm{OBB}}-1}, \quad \nu \geq\nu_\mathrm{break}, \\
        &(\frac{R_\mathrm{OBB}}{D})^2\pi\frac{2h\nu_\mathrm{break}^3}{c^2} \frac{1}{e^{h\nu_\mathrm{break}/k_\mathrm{B} T_\mathrm{OBB}}-1}\\
        &\quad \quad \quad \times (\nu/\nu_\mathrm{break})^{(4s-6)/(3s-2)}, \quad \nu \leq \nu_\mathrm{break}.
    \end{aligned}
    \right
.
\label{eq9}
\end{equation}
Since Eq.~\ref{eq4} depends only weakly on temperature, we adopt $T=T_\mathrm{OBB}$. The density slope $s$ and the break frequency $\nu_\mathrm{break}$ are treated as free parameters.

Using Eqs.~\ref{eq5} and \ref{eq6}, the normalization $K(\rho_0,s,T)$ can be expressed as 
\begin{equation}
\begin{aligned}
    K(\rho_0,s,T)=&(\frac{R_\mathrm{OBB}}{D})^2\pi B_{\nu_\mathrm{break}}(T_\mathrm{OBB})\\
    & \times 4\pi D^2/\nu_\mathrm{break}^\frac{4s-6}{3s-2},
\end{aligned}
\label{eq10}
\end{equation}
which allows the density normalization $\rho_0$ to be determined through Eq.~\ref{eq7}. Combining the fitted values of $\nu_\mathrm{break}$ and $s$ with Eqs.~\ref{eq2} and \ref{eq7}, we further derive the thermalization radius $r_\mathrm{th,\nu_\mathrm{break}}$ and the density at that radius, which are also presented in Table.~\ref{tab2}. The best-fitting models are plotted in Fig.~\ref{fig7}.

\begin{table}
	\centering
	\caption{Best-fitting values of the density-profile index $s$, the break frequency $\nu_\mathrm{break}$, the thermalization radius $r_\mathrm{th,\nu_\mathrm{break}}$, density normalization $\rho_0$, the thermalization radius of the $K$-band photons $r_\mathrm{th,K}$, and the total mass enclosed within this radius, $M_\mathrm{r_\mathrm{th,K}}$, for the three epoch with NIR observation. The UV--optical blackbody radius, $R_\mathrm{OBB}$, is also listed for comparison. For epochs 1 and 2, the fits used both the $J$- and $K$-band measurements. For epoch 3, we use the $J$ measurement supplemented with the $K$-band observation from epoch 2.}
	\label{tab2}
    \setlength{\tabcolsep}{8pt}
	\begin{tabular}{cccc} 
		\hline
        Epoch & 1 & 2 & 3 \\
        \hline
        Time & +41~d & +68~d & +101~d \\
        \hline
        NIR observations & $J$, $K$ & $J$, $K$ & $J$, Epoch 2 $K$ \\
        \hline
        $s$ & $1.76^{+0.35}_{-0.30}$ & $1.39^{+0.34}_{-0.23}$ & $1.58^{+0.45}_{-0.23}$ \\
        $\nu_\mathrm{break}~(\times10^{14}\mathrm{Hz})$ & $2.89^{+0.24}_{-0.33}$ & $2.63^{+0.27}_{-0.37}$ &  $3.67^{+0.76}_{-0.44}$ \\
        $r_\mathrm{th,\nu_\mathrm{break}}~(\mathrm{\times10^{14}~cm})$ & $10.88^{+3.01}_{-3.02}$ & $4.37^{+2.13}_{-1.94}$ &  $4.95^{+2.22}_{-1.39}$ \\
        $\rho_0~(\mathrm{\times10^{-14}~g~cm^{-3}})$ & $6.37^{+3.37}_{-1.93}$ & $2.91^{+0.63}_{-0.52}$ &  $3.19^{+1.16}_{-0.44}$\\
        $r_\mathrm{th,K}~(\mathrm{\times10^{15}~cm})$ & $1.64^{+0.32}_{-0.39}$ & $0.75^{+0.24}_{-0.31}$ &  $0.96^{+0.26}_{-0.19}$ \\
        $M_\mathrm{r_\mathrm{th,K}}~(M_\odot)$ & $0.25^{+0.13}_{-0.12}$ & $0.04\pm0.03$ & $0.06^{+0.05}_{-0.03}$\\
        $R_\mathrm{OBB}~(\mathrm{\times10^{14}~cm})$ & $5.70^{+0.31}_{-0.30}$ & $3.88\pm0.21$ & $3.53^{+0.35}_{-0.30}$\\
		\hline
	\end{tabular}
\end{table}

We then use Eq.~\ref{eq9} to fit the three combined UV--optical--NIR spectra constructed with data taken closely in time (see Table~\ref{tab2} for more details). Because the source was not detected in the $K$ band during epoch 3, the NIR component cannot be well constrained by the $J$-band measurement alone. We therefore performed a fit in which the epoch 2 $K$-band measurement was adopted for epoch 3, since the $J$-band flux keeps constant in epoch 2 and 3, and we do not expect large variation in $K$ band. The best-fitting parameters obtained both with this additional $K$-band constraint are reported in Table~\ref{tab2}. 

Overall, we find $\nu_\mathrm{break}$ systematically increases with time, while $s$ remains consistent within the uncertainties, suggesting that the overall density structure of the reprocessing medium remains approximately unchanged. Because $\nu_\mathrm{break}$ marks the transition between the UV--optical blackbody component and the low-frequency excess, $r_\mathrm{th,\nu_\mathrm{break}}$ represents the innermost thermalization radius associated with the frequency-dependent reprocessing component. It is therefore expected to lie outside the optical photosphere, as is indeed the case at all three epochs. Both $r_{\mathrm{th},\nu_\mathrm{break}}$ and the optical photospheric radius decrease with time. Meanwhile, the inferred density normalization, $\rho_0$, declines modestly, consistent with a gradual decrease in the density of the reprocessing or outflowing material.

To estimate the mass of the outflowing material, we calculate the thermalization radius at the $K$-band frequency,
$r_\mathrm{th,K}$, which traces a relatively outer region of the reprocessing layer.
Integrating the density profile within this radius yields $r_\mathrm{th,K}\sim10^{15}\ \mathrm{cm}$ and an enclosed mass, $M_\mathrm{K}$, ranging from several $10^{-2}$ to several $10^{-1}~M_\odot$. Both quantities initially decrease and subsequently remain approximately constant. The decline in $M_\mathrm{K}$ is driven primarily by the inward evolution of the thermalization radius and should therefore not be interpreted as a direct measurement of the temporal evolution of the total outflow mass.

A caveat of this estimate is that the disrupted stellar mass inferred from MOSFiT is only $\sim$$0.3~M_\odot$, comparable to the inferred lower limit on $M_\mathrm{K}$. This apparent tension likely reflects the simplifying assumptions of the free--free emission model, particularly the adoption of a homogeneous spherical medium. If the reprocessing material is clumpy, anisotropic, or occupies only a fraction of the available volume, the true gas mass could be substantially smaller than the value inferred under the spherical approximation. Consequently, the estimated mass should be viewed primarily as an order-of-magnitude indication of a dense, extended reprocessing layer rather than a precise measurement of the total outflow mass.

\begin{figure}
    \centering
    \includegraphics[width=0.7\columnwidth]{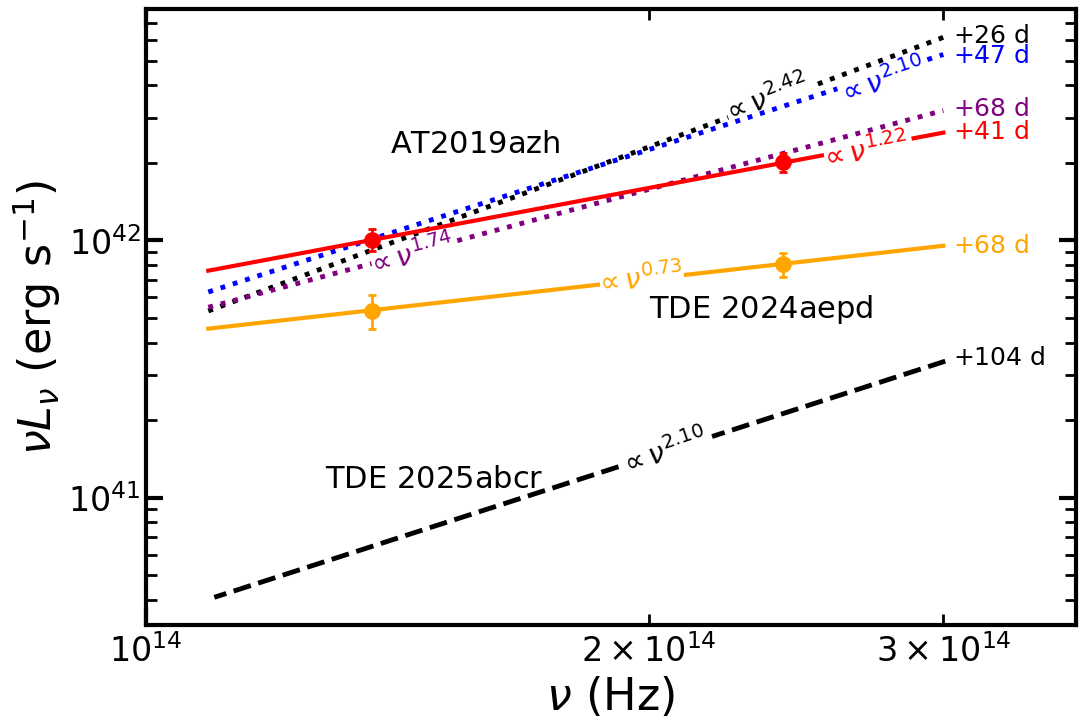}
    \caption{Comparison of the three TDEs with reported NIR excesses to date. The colored solid, dotted, and dashed lines represent the best-fitting power-law models for 2024aepd, AT2019azh \citep{2026A&A...708A.139R}, and TDE~2025abcr \citep{patra2026jwstkeckobservationsoffnuclear}, respectively. The labeled time offsets are measured relative to the discovery date of each event.}
    \label{fig8}
\end{figure} 

In Fig.~\ref{fig8}, we compare the best-fitting power-law models for the NIR emission of AT2019azh and TDE~2025abcr with those of 2024aepd. The NIR spectra of 2024aepd appear flatter than those inferred for the other two events. Within the frequency-dependent thermalization framework, this flatter NIR spectrum implies a shallower density profile in the extended reprocessing layer.
Moreover, whereas the NIR spectral slope of AT2019azh evolves significantly and becomes progressively flatter with time, $\alpha_\mathrm{IR}$ in 2024aepd remains consistent within the uncertainties. This may indicate substantial evolution in the density structure of the extended material in AT2019azh, but a comparatively stable structure in 2024aepd.
The physical origin of these differences remains unclear and warrants further investigation. More generally, if the NIR excess arises from free--free emission in an extended outflowing layer, similar excesses may be common among TDEs. However, the current sample is too limited to determine whether such emission is ubiquitous.

Interestingly, the recently discovered Little Red Dots (LRDs) exhibit V-shaped UV--optical SEDs broadly resembling that of 2024aepd, although the spectral turnover in LRDs generally occurs near the Balmer limit at $3646\,\text{\AA}$ (e.g., \citealt{Setton2025,2025MNRAS.544.3900J,Zhang2026}). One proposed interpretation invokes an accreting black hole, potentially undergoing super-Eddington accretion, embedded in dense, optically thick gas \citep{2025ApJ...980L..27I,2025ApJ...994..113L}. In this scenario, the large bound--free opacity of hydrogen in the $n=2$ state suppresses the continuum blueward of the Balmer limit, while emission from the dense gas produces the red rest-frame optical continuum \citep{2025ApJ...980L..27I,2025MNRAS.544.3900J}. The observed blue UV component may instead originate from surrounding young stars or from leaked or scattered AGN emission.

It has also been proposed that some LRDs may be associated with TDEs occurring in dense stellar systems at high redshift \citep{2025ApJ...984L..55B}. However, explaining a substantial fraction of the LRD population through this scenario may require an enhanced TDE rate or a long-lived phase powered by recurrent tidal disruptions. Based on the discussion presented here, if some LRDs are indeed powered by TDEs, their luminosities and spectral-break wavelengths may evolve with time. Such evolution could be tested through long-term monitoring or by searching for a luminosity dependence of the turnover wavelength across the population.

\subsection{$\Gamma$--$L_{\mathrm{X}}$ relationship in TDEs}
Around day 178, corresponding to $\sim150$~days after optical peak, the X-ray spectrum of 2024aepd transitions from a blackbody-dominated state to a pure power-law state. The nearly constant luminosity of the \texttt{powerlaw} component throughout the period studied may reflect changes in coronal properties, such as the electron temperature and optical depth. We disfavor a faint AGN contribution because neither the host-galaxy SED modeling nor the pre-TDE optical spectrum provides evidence for persistent AGN activity.
This evolution therefore suggests a spectral-state transition occurring at an Eddington ratio of $L_{\rm X}/L_{\rm Edd}\sim0.2\%$. Analogous transitions from a disk-dominated high/soft state to a corona-dominated low/hard state are well established in XRBs, where they typically occur at $L_{\rm X}/L_{\rm Edd}\sim1$--$2\%$ \citep{2006ARA&A..44...49R}.

Transitions from thermal-dominated to non-thermal-dominated X-ray spectra have been increasingly reported in TDEs. However, the inferred transition times span a remarkably broad range, from approximately one hundred to several hundreds of days (e.g., ASASSN-15oi, \citealt{2025ApJ...983...29H}; AT2018fyk, \citealt{2021ApJ...912..151W}; AT2020ocn, \citealt{2024ApJ...966..160G}; AT2021ehb, \citealt{2022ApJ...937....8Y}).
Recently, \citet{2026arXiv260221624Y} proposed an anti-correlation between the transition time, defined as the epoch when $L_{\rm disk}=3\%\,L_{\rm Edd}$, and black hole mass, suggesting that state transitions occur more rapidly in systems hosting more massive black holes. They further interpreted the non‑thermal‑dominated spectra observed from the earliest epochs in AT2024lhc and AT2024kmq as a consequence of their initially low accretion rates, given the inferred black hole masses of $10^8~M_\odot$.
This trend is qualitatively expected, as lower-mass black holes can sustain super-critical or near-critical accretion rates for longer periods than their higher-mass counterparts.
Nevertheless, the observed diversity of transition times suggests that black hole mass alone is unlikely to determine the evolution. For example, AT2018fyk, which hosts a black hole approximately an order of magnitude more massive than those inferred for AT2020ocn, AT2021ehb, and 2024aepd, exhibits a comparable transition timescale. This indicates that additional parameters likely contribute to regulating the transition, including the mass of the disrupted star, the penetration factor of the encounter, and the efficiency of debris circularization and disk formation.

State transitions in accreting black hole systems are often accompanied by characteristic changes in the relationship between the photon index, $\Gamma$, and the X-ray luminosity, $L_{\rm X}$. In XRBs, $\Gamma$ is positively correlated with $L_{\rm X}$ at relatively high accretion rates ($L_{\rm X}/L_{\rm Edd}\gtrsim$ a few percent), whereas the correlation reverses and becomes negative at lower accretion rates \citep[e.g.,][]{2008ApJ...682..212W,2015MNRAS.447.1692Y}. Similar behavior has been observed in AGN, where luminous sources generally show a positive $\Gamma$--$L_{\rm X}$ correlation \citep{2004ApJ...607L.107W}, while low-luminosity AGN exhibit an anti-correlation \citep{2009MNRAS.399..349G}. The transition between the two regimes occurs near a bolometric luminosity of $\sim$$1\%L_{\rm Edd}$ \citep{2009ApJ...705.1336C}.

These trends are commonly interpreted within models in which the structure of the inner accretion flow changes with accretion rate, such as the truncated-disk scenario \citep{2011MNRAS.417..280S,2011MNRAS.414.3330V} and the disk--corona evaporation framework \citep{2006A&A...454L...9L,2013ApJ...764....2Q,2013ApJ...777..102Q}.
In the truncated-disk model, the geometrically thin disk is truncated at a certain radius, with a radiatively inefficient accretion flow (RIAF) occupying the inner region. At low Eddington ratios, the dominant seed photons for Comptonization are produced internally through synchrotron and bremsstrahlung emission in the RIAF. As the accretion rate increases, the optical depth of the hot flow rises, enhancing Comptonization and producing a harder X-ray spectrum. This behavior gives rise to the negative $\Gamma$--$L_{\rm X}$ correlation. At higher Eddington ratios, the truncation radius moves inward as the accretion rate increases, allowing the thin disk to extend closer to the black hole. The resulting increase in the disk seed-photon flux strengthens Compton cooling of the hot inner flow, lowers the electron temperature, and produces a softer X-ray spectrum. This transition naturally accounts for the positive $\Gamma$--$L_{\rm X}$ correlation. A similar interpretation applies in the disk--corona evaporation model, while the positive $\Gamma$--$L_{\rm X}$ correlation can also be explained by the disk--corona frameworks \citep[e.g.,][]{2009MNRAS.394..207C,2025MNRAS.544.1748X}.

\begin{figure}
    \centering
    \includegraphics[width=0.7\textwidth]{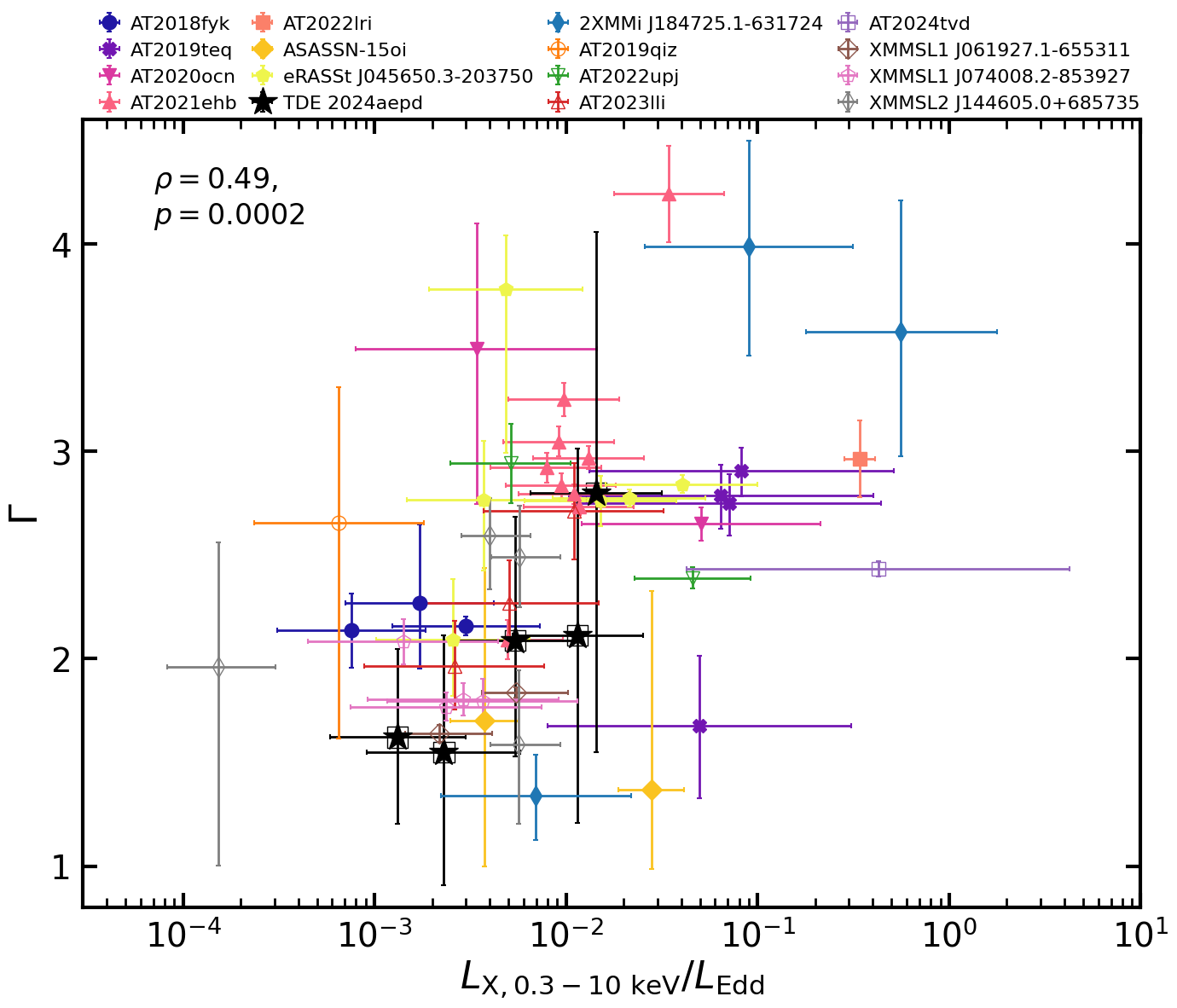}
    \caption{Photon index $\Gamma$ as a function of Eddington ratio for TDEs exhibiting a hard X-ray excess. In addition to 2024aepd, the sample includes XMMSL2~J144605.0+685735 \citep{2019A&A...630A..98S}, 2XMMi~J184725.1--631724 \citep{2011ApJ...738...52L},  XMMSL1~J061927.1--655311 \citep{2014A&A...572A...1S}, XMMSL1~J074008.2--853927 \citep{2017A&A...598A..29S}, ASASSN-15oi \citep{2016MNRAS.463.3813H}, AT2018fyk \citep{2021ApJ...912..151W}, AT2019qiz \citep{2020MNRAS.499..482N}, AT2019teq \citep{2026ApJ...999..265B}, AT2020ocn \citep{2024ApJ...966..160G}, eRASSt~J045650.3--203750 \citep{2023A&A...669A..75L}, AT2021ehb \citep{2022ApJ...937....8Y}, AT2022lri \citep{2024ApJ...976...34Y}, AT2022upj \citep{2024ApJ...977..258N}, AT2023lli \citep{2024ApJ...964L..22H}, 
    and AT2024tvd \citep{2025ApJ...992L..18S}. The Spearman's rank correlation coefficient, $\rho$, and the corresponding $p$-value are also shown in the figure.}
    \label{fig9}
\end{figure}

To investigate the $\Gamma$--$L_{\mathrm{X}}$ relationship in TDEs, we select all X-ray TDEs from the Open mulTiwavelength Transient Event Repository \citep[OTTER;][]{2026JOSS...11.9516F,2026ApJ...999..243F} that have been reported to exhibit a hard X-ray excess in addition to a thermal component in their XMM-Newton spectra.
After excluding observations for which the non-thermal component has $\Gamma>5$, as such unusually steep spectra are unlikely to arise from Comptonization, we obtain a final sample of 15 events.
The use of XMM-Newton/EPIC-pn data provides the advantages of a large effective area and relatively long exposures, enabling any non-thermal component to be more reliably separated from the thermal emission.
All spectra are grouped to contain at least 25 counts per bin, and $\chi^2$ statistics are adopted for spectral fitting. For each observation, we fit the spectrum over the energy range from 0.3~keV to the highest energy at which the net source count rate remained above the background level. With the exception of AT2022lri, for which an additional \texttt{gaussian} component is required in two epochs, all spectra are modeled using \texttt{TBabs*zashift*simpl*diskbb}. We then derive the corresponding 0.3--10~keV X-ray luminosities from the best-fitting models.

The derived $\Gamma$--$L_{\mathrm{X}}/L_{\mathrm{Edd}}$ relation is shown in Fig.~\ref{fig9}, together with our best-fitting results for 2024aepd. To assess whether $\Gamma$ and $L_{\mathrm{X}}/L_{\mathrm{Edd}}$ exhibit a monotonic correlation without assuming a specific functional form, we calculate the Spearman rank correlation coefficient. We obtain $\rho=0.49$ with $p=0.0002$, indicating a statistically significant positive correlation across the full sample. 
Restricting the analysis to the optically selected subsample yields a weaker correlation, with $\rho=0.32$ and $p=0.07$. Unlike the classical $\Gamma$--$L_{\mathrm{X}}/L_{\mathrm{Edd}}$ relation observed in other accreting systems, no clear anti-correlation branch is identified in our TDE sample. Although 2024aepd undergoes a state transition at around 0.2\% Eddington ratio, the large uncertainties and the limited number of observations at low Eddington ratios prevent us from statistically identifying a turnover in the relation. Extending TDE samples to lower luminosities will therefore be essential for determining whether such a turnover exists.

It should be noted that our sample includes only TDEs listed in OTTER for which a non-thermal X-ray component has been reported and XMM-Newton/EPIC-pn observations are available. TDEs exhibiting hard-to-soft spectral transitions may therefore be absent from our sample, including AT2018hyz \citep{2020MNRAS.497.1925G} and AT2019avd \citep{Wang2024}. Moreover, we require each spectrum to contain at least five spectral bins, ensuring that the number of degrees of freedom is greater than or equal to one. This criterion may exclude epochs during which the source is particularly faint. Although spectra from such observations would likely be poorly constrained, this selection may introduce a bias against identifying the turnover.

\section{Conclusion}
\label{sect:conclusion}
We present multi-wavelength observations of 2024aepd spanning the NIR to soft X-rays and summarize our main results as follows:

\begin{enumerate}[leftmargin=2em, label=\textbullet, itemsep=0.4em, topsep=0.4em]
\item The blackbody temperature, radius and the luminosity of the UV--optical emission exhibit significant decrease throughout the observations. The broad emission lines fade on a shorter timescale than the continuum.

\item An early-time NIR excess above the UV--optical blackbody component is detected. Free--free emission from an extended reprocessing layer provides a more plausible explanation for this excess, whereas a conventional dust-echo origin is less favored because the light-travel time to the inferred sublimation radius is substantially longer than the observed lag between the optical and NIR peaks. 

\item The X-ray spectrum exhibits a hard excess above the dominant thermal component from the earliest observations and is well described by a combination of thermal and non-thermal emission.

\item As the X-ray luminosity declines, the hardness ratio increases and the photon index decreases. At approximately $175$~days, the non-thermal component becomes dominant, indicating a transition from a soft to a hard X-ray state. We interpret this transition as the strengthening and eventual dominance of a corona, analogous to state transitions observed in XRBs and commonly invoked in TDEs.

\item A positive correlation between $\Gamma$ and $L_{\rm X}$ is identified in the compiled sample of X-ray TDEs. The absence of the negative branch observed in low-accretion-rate XRBs and AGN may be due to the lack of sufficiently low-luminosity coverage in the current sample. 

\end{enumerate}

Overall, our results provide evidence for frequency-dependent reprocessing at early times and evolving radiative-transfer conditions in TDEs, placing new constraints on the structure of the reprocessing layer and the formation of disk--corona systems around supermassive black holes. Similar early-time NIR excesses may be common among TDEs, motivating systematic NIR follow-up of a larger sample at early epochs.

\begin{acknowledgements}
We thank the referee for the valuable and constructive comments, which have helped improve the manuscript.
We thank the participants of the TDE FORUM (Full-process Orbital to Radiative Unified Modeling) online seminar series for their inspiring discussions. We also thank Lixin Dai, Lars L. Thomsen, and Tinggui Wang for the useful discussions.
This research was supported by the National Natural Science Foundation of China (NSFC) under grant No. 12588202, the New Cornerstone Science Foundation through the New Cornerstone Investigator Program and the XPLORER PRIZE, and the Strategic Priority Program of the Chinese Academy of Sciences under grant No. XDB0550203. 
T.M.R. is a member of the Cosmic Dawn Center (DAWN), which is funded by the Danish National Research Foundation under grant DNRF140. T.M.R. acknowledges support from the Research Council of Finland project 350458. 
S.Z. is supported by the National Key Research and Development Program of China (No. 2024YFA1611603) and the Yunnan Key Laboratory of Survey Science (No.202449CE340002). I.A. acknowledges support from the European Research Council (ERC) under the European Union’s Horizon 2020 research and innovation program (grant agreement number 852097), from the Israel Science Foundation (grant number 2752/19), from the United States - Israel Binational Science Foundation (BSF; grant number 2024812), and from the Pazy foundation (grant number 216312).
This project has received funding from the European Research Council (ERC) under the European Union’s Horizon Europe research and innovation programme (grant agreement No. 101221278, project OUTLIERS).
D.A. also acknowledges financial support from the Spanish Ministry of Science and Innovation (MICINN) under the 2021 Ram\'on y Cajal program MICINN RYC2021-032609.
P.C. acknowledges financial support from the Secretary of Universities and Research (Government of Catalonia) and by the Horizon 2020 Research and Innovation Programme of the European Union under the Marie Sk\l{}odowska-Curie and the Beatriu de Pin\'os 2024 BP 00125 programme, and from the Centro Superior de Investigaciones Cient\'ificas (CSIC) under the Spanish program Unidad de Excelencia Mar\'ia de Maeztu CEX2020-001058-M, financed by MCIN/AEI/10.13039/501100011033, and by the MaX-CSIC Excellence Award MaX4-SOMMA-ICE and support via Research Council of Finland (grant 340613).
F.P. acknowledges support from the Spanish Ministerio de Ciencia, Innovaci\'on y Universidades (MICINN) under grant numbers PID2022-141915NB-C21.
H.Z. and N.L. acknowledge the support from the National Natural Science Foundation of China (NSFC; grant Nos. 12120101003 and 12233008) and the Programs of National Astronomical Observatories Chinese Academy of Sciences (Grant Nos. E5ZQ7801, E5ZB7801, and E4TG2001).
T.E.M.B. is funded by Horizon Europe ERC grant no. 101125877. S.M. acknowledges financial support from the Research Council of Finland project 350458.
C.P.G. acknowledges financial support from grant RYC2024-050959-I, funded by MICIU/AEI/10.13039/501100011033 and the FSE+, as well as from projects PID2023-151307NB-I00, PIE 20215AT016, and CEX2020-001058-M, and the MaX-CSIC Excellence Award MaX4-SOMMA-ICE.
B.K. acknowledges the support from the ``Special Project for High-End Foreign Experts", Xingdian Funding from Yunnan Province, and the National Key Research and Development Program of China (2024YFA1611603). 
F.O. acknowledges support from the INAF-Large Grant 2024:"Envisioning Tomorrow: prospects and challenges for multimessenger astronomy in the era of Rubin and Einstein Telescope"; the INAF-GO Large Grant: "Exploitation of optical and near-infrared followup data of Gamma-ray Bursts" and the INAF-MINIGRANT (2023): "SeaTiDE - Searching for Tidal Disruption Events with ZTF: the Tidal Disruption Event population in the era of wide field surveys".   

This study used observations collected at the European Organisation for Astronomical Research in the Southern Hemisphere, Chile, as part of ePESSTO+ (the advanced Public ESO Spectroscopic Survey for Transient Objects Survey). ePESSTO+ observations were obtained under ESO program ID 112.25JQ.
The data presented here were obtained in part with ALFOSC, which is provided by the Instituto de Astrofisica de Andalucia (IAA) under a joint agreement with the University of Copenhagen and NOT.

This research used data obtained with the Dark Energy Spectroscopic Instrument (DESI). DESI construction and operations is managed by the Lawrence Berkeley National Laboratory. This material is based upon work supported by the U.S. Department of Energy, Office of Science, Office of High-Energy Physics, under Contract No. DE–AC02–05CH11231, and by the National Energy Research Scientific Computing Center, a DOE Office of Science User Facility under the same contract. Additional support for DESI was provided by the U.S. National Science Foundation (NSF), Division of Astronomical Sciences under Contract No. AST-0950945 to the NSF’s National Optical-Infrared Astronomy Research Laboratory; the Science and Technology Facilities Council of the United Kingdom; the Gordon and Betty Moore Foundation; the Heising-Simons Foundation; the French Alternative Energies and Atomic Energy Commission (CEA); the National Council of Science and Technology of Mexico (CONACYT); the Ministry of Science and Innovation of Spain (MICINN), and by the DESI Member Institutions: www.desi.lbl.gov/collaborating-institutions. The DESI collaboration is honored to be permitted to conduct scientific research on Iolkam Du’ag (Kitt Peak), a mountain with particular significance to the Tohono O’odham Nation. Any opinions, findings, and conclusions or recommendations expressed in this material are those of the author(s) and do not necessarily reflect the views of the U.S. National Science Foundation, the U.S. Department of Energy, or any of the listed funding agencies. The SYSU 80 cm infrared telescope is operated and managed by the Department of Astronomy, Sun Yat-sen University.

This research was partly based on observations made with the Nordic Optical Telescope (program IDs: P59-506 \& P68-505) owned in collaboration by the University of Turku and Aarhus University, and operated jointly by Aarhus University, the University of Turku and the University of Oslo, representing Denmark, Finland and Norway, the University of Iceland and Stockholm University at the Observatorio del Roque de los Muchachos, La Palma, Spain, of the Instituto de Astrofisica de Canarias. This publication makes use of data products from the Two Micron All Sky Survey, which is a joint project of the University of Massachusetts and the Infrared Processing and Analysis Center/California Institute of Technology, funded by the National Aeronautics and Space Administration and the National Science Foundation. 

The Multi-channel Photometric Survey Telescope (Mephisto) is developed at and operated by the South-Western Institute for Astronomy Research of Yunnan University (SWIFAR-YNU), funded by the ``Yunnan University Development Plan for World-Class University" and ``Yunnan University Development Plan for World-Class Astronomy Discipline". The Mephisto team acknowledges support from the Key Laboratory of Survey Science of Yunnan Province (202449CE340002), the ``Science \& Technology Champion Project'' (202005AB160002), and from two ``Team Projects" -- the ``Top Team'' (202305AT350002) and the ``Innovation Team'' (202105AE160021), funded by the ``Yunnan Revitalization Talent Support Program". 

The Australia Telescope Compact Array is part of the Australia Telescope National Facility (grid.421683.a) which is funded by the Australian Government for operation as a National Facility managed by CSIRO. We acknowledge the Gomeroi people as the traditional owners of the Observatory site. e-MERLIN is a National Facility operated by the University of Manchester at Jodrell Bank Observatory on behalf of STFC, part of UK Research and Innovation.

\end{acknowledgements}

\appendix                  

\section{MOSFiT fitting results and host galaxy SED}
\label{sect:AppendixA}
\begin{figure}
    \centering
    \includegraphics[width=0.6\columnwidth]{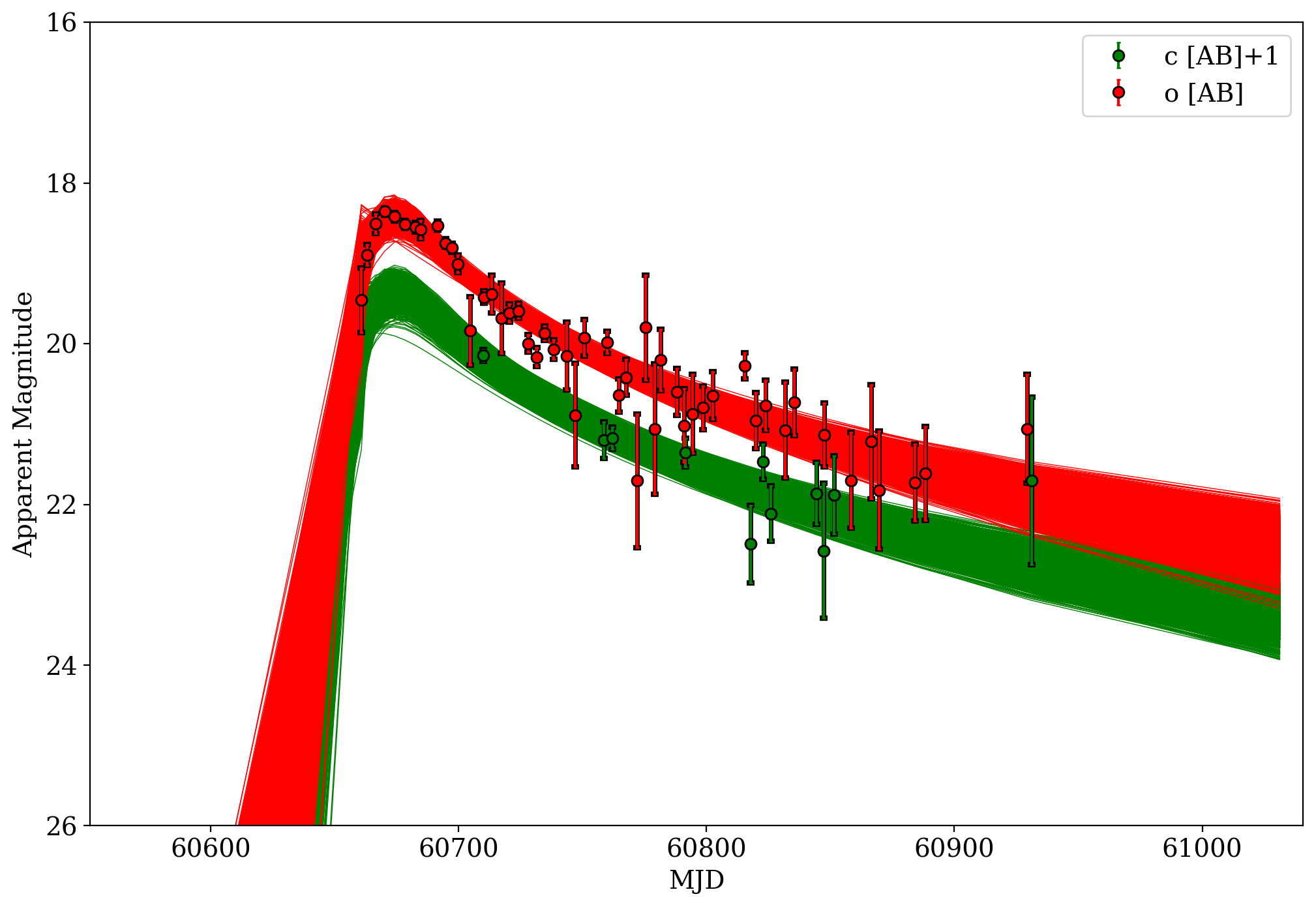}
    \caption{MOSFiT light-curve fitting results for the ATLAS optical photometry of 2024aepd.}
    \label{figA1}
\end{figure}

\begin{figure}
    \centering
    \includegraphics[width=0.6\columnwidth]{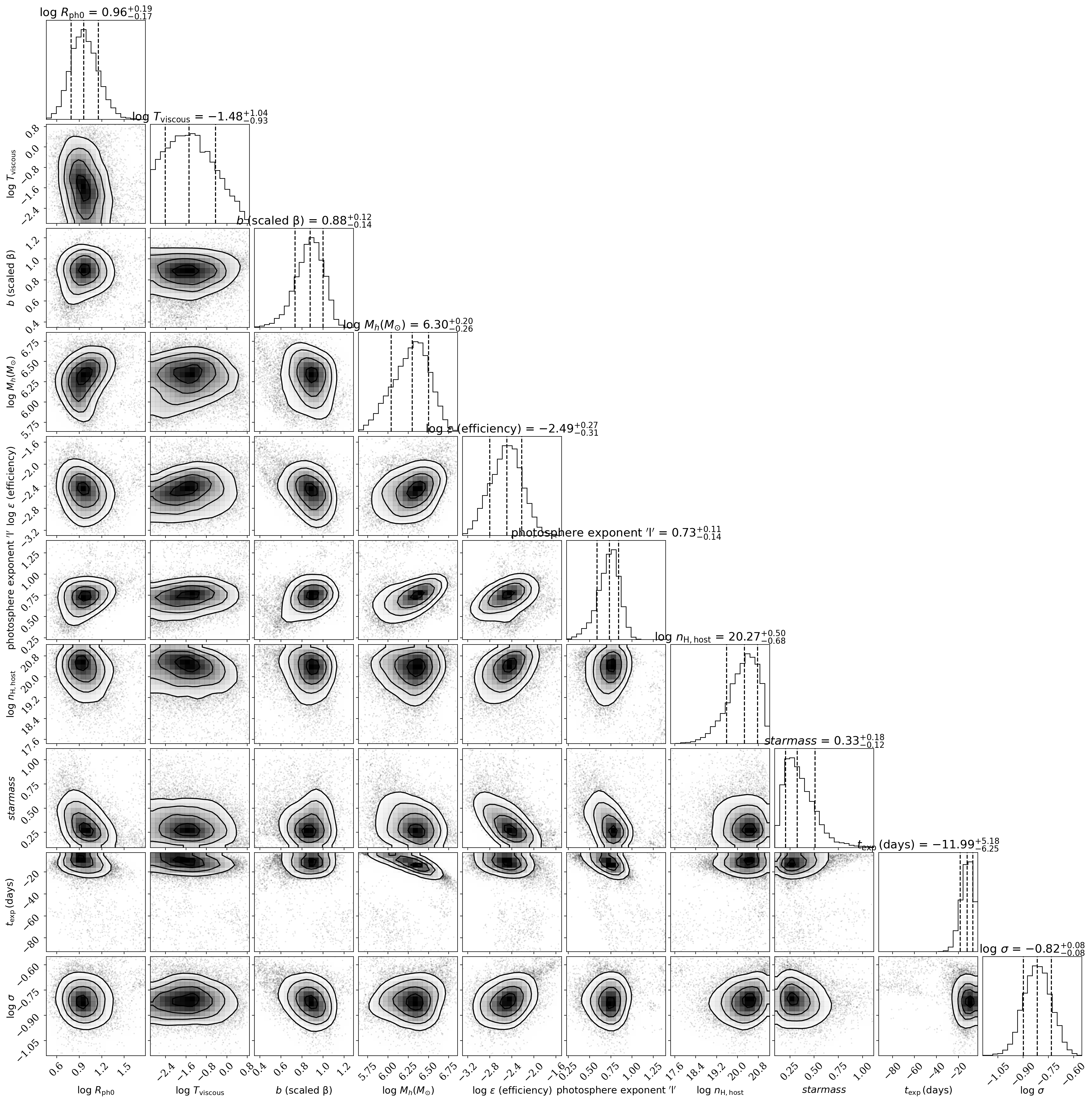}
    \caption{Corner plots of the parameters inferred from the MOSFiT light-curve fitting using nested sampling.}
    \label{figA2}
\end{figure}

\begin{figure}
    \centering
    \includegraphics[width=0.6\columnwidth]{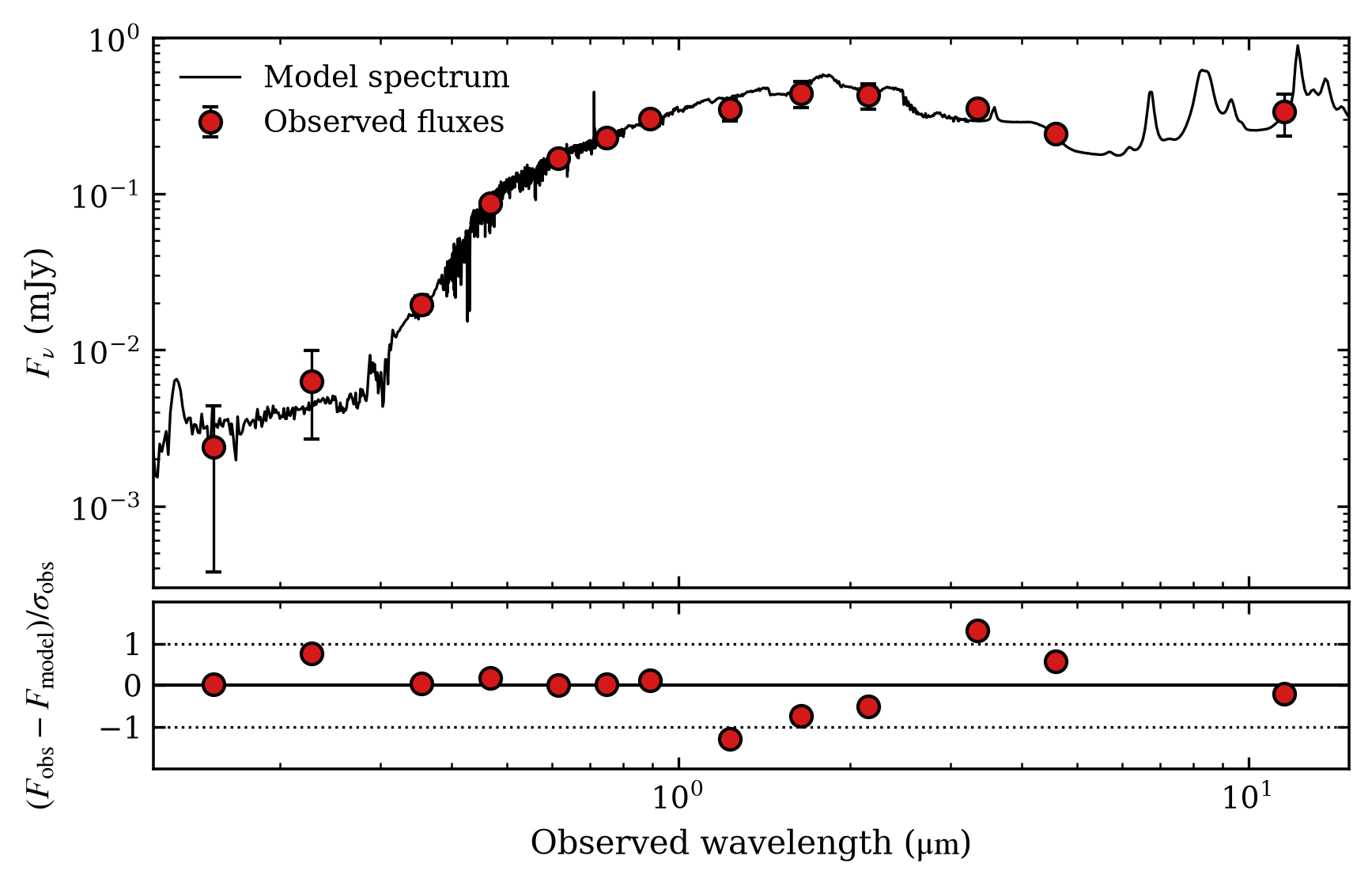}
    \caption{Host-galaxy SED fitted with \texttt{Prospector}. The red circles show the observed photometry, and the black solid curve represents the best-fitting total SED. The lower panel presents the relative residuals.}
    \label{fig:host_sed}
\end{figure}

We fit the ATLAS $c$- and $o$-bands light curve using the TDE model implemented in MOSFiT. The best-fitting parameters and model light curves are presented in Table~\ref{tabA1}, Fig.~\ref{figA1} and Fig.~\ref{figA2}. We model the host-galaxy SED using \texttt{Prospector}, and show the best-fitting result in Fig.~\ref{fig:host_sed}.
\begin{table}
	\centering
	\caption{MOSFiT TDE model parameter fits}
	\label{tabA1}
	\begin{tabular}{lccc} 
		\hline
        Quantity & Value & Units \\
		\hline
  		$\mathrm{log}~R_\mathrm{ph0}$~(photosphere power-law constant) & $0.96^{+0.19}_{-0.17}$ & - \\      
		$\mathrm{log}~T_\mathrm{viscous}$~(viscous delay time-scale) & $-1.48^{+1.04}_{-0.93}$ & d \\
		$b$~(scaled impact parameter) & $0.88^{+0.12}_{-0.14}$ & -\\
        $\mathrm{log}~M_\mathrm{h}$~(SMBH mass) & $6.30^{+0.20}_{-0.26}$ & $M_\odot$ \\
        $\mathrm{log}~\epsilon$~(efficiency) & $-2.49^{+0.27}_{-0.31}$ & - \\
        $l$~(photosphere power-law exponent) & $0.73^{+0.11}_{-0.14}$ & - \\
        $\mathrm{log}~n_\mathrm{H,host}$~(local hydrogen column density) & $20.27^{+0.50}_{-0.68}$ & $\mathrm{cm^{-2}}$ \\
        $M_*$~(stellar mass) & $0.33^{+0.18}_{-0.12}$ & $M_\odot$ \\
        $t_\mathrm{exp}$~(time of disruption) & $-11.99^{+5.18}_{-6.25}$ & d \\
        $\mathrm{log}~\sigma$~(model variance) & $-0.82\pm 0.08$ & - \\
		\hline
	\end{tabular}
\end{table}

\section{Best-fit results for the combined X-ray spectra using different spectral models}
\begin{figure}
    \centering
    \includegraphics[angle=-90,width=0.6\columnwidth]{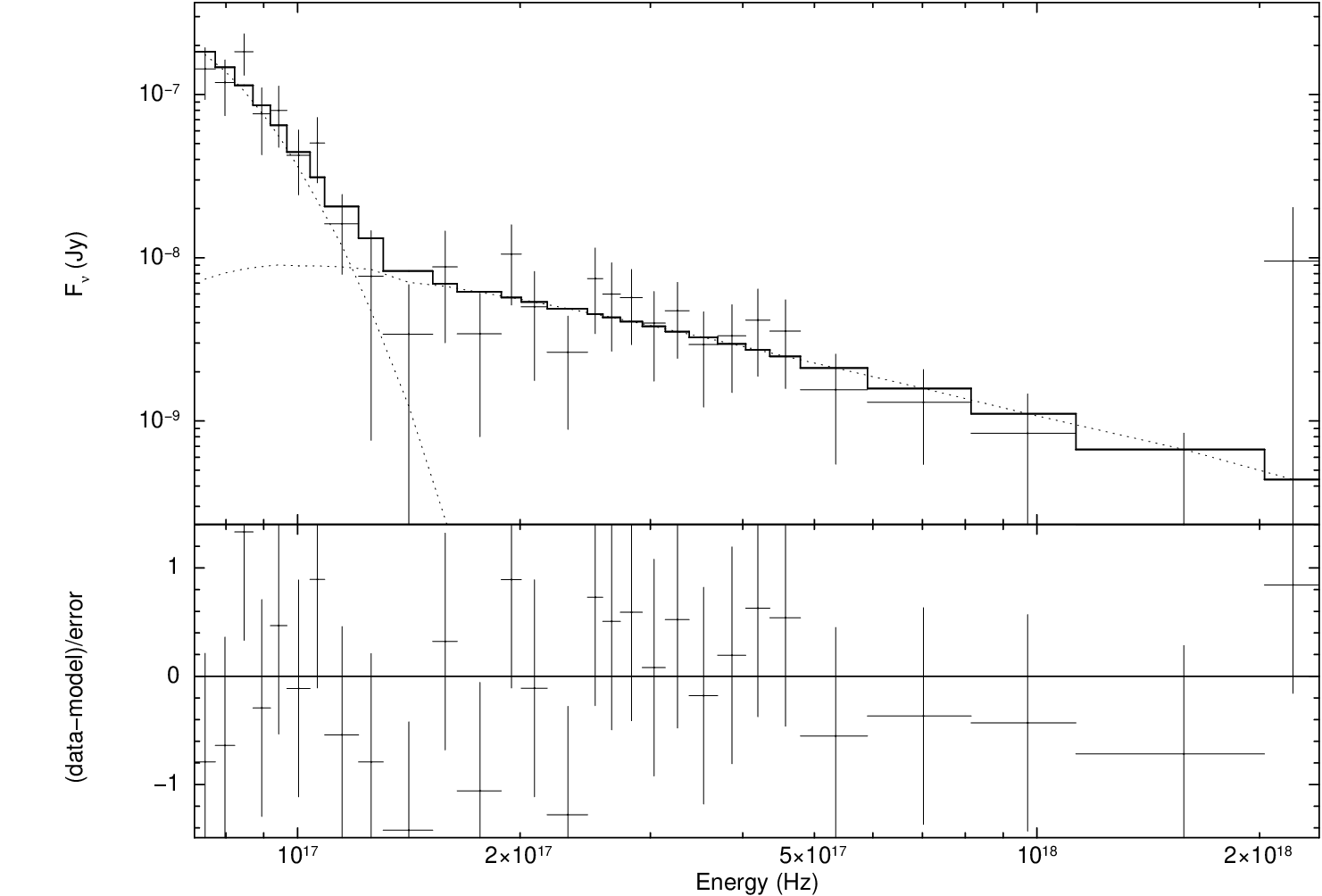}
    \caption{The best-fit results of the combined X-ray spectrum with the \texttt{tbabs*zashift(diskbb+powerlaw)} model.}
    \label{figB1}
\end{figure}

We fit the combined X-ray spectrum with the \texttt{diskbb+powerlaw} model and several alternative combinations of thermal and non-thermal components. The best-fit results are presented in Fig.~\ref{figB1} and Table~\ref{tabB1}.

\begin{table}
	\centering
	\caption{Best-fitting parameters for the combined X-ray spectra obtained using different combinations of thermal and non-thermal models. $F_\mathrm{diskbb}$, $F_\mathrm{tdediscspec}$ and $F_\mathrm{pl}$ are the unabsorbed flux of the \texttt{diskbb}, \texttt{tdediscspec} and the non-thermal component integrated from 0.3 to 10~keV. $f_\mathrm{sc}$ denotes the fraction of seed photons that are scattered into the non-thermal component.}
	\label{tabB1}
	\begin{tabular}{lccc} 
		\hline
		   & \texttt{diskbb+powerlaw} & \texttt{simpl*diskbb}  \\
		\hline
  		$T_\mathrm{in}~\mathrm{(keV)}$ & $0.043\pm0.007$ & $0.043\pm0.007$\\      
		$F_\mathrm{diskbb}~\mathrm{(erg~cm^{-2}~s^{-1})}$ & $1.15^{+0.22}_{-0.24} \times 10^{-13}$ & $1.15^{+0.22}_{-0.24} \times 10^{-13}$\\
		$\Gamma$ & $2.07^{+0.27}_{-0.26}$ & $2.07^{+0.27}_{-0.26}$\\
        $f_\mathrm{sc}$ & - & $2.14^{+3.43}_{-1.38} \times 10^{-3}$\\
        $F_\mathrm{pl}~\mathrm{(erg~cm^{-2}~s^{-1})}$ & $4.79^{+0.65}_{-0.69} \times 10^{-14}$ & $4.76^{+3.17}_{-3.38} \times 10^{-14}$\\
		\hline
          & \texttt{tdediscspec+powerlaw} & \texttt{simpl*tdediscspec}\\
        \hline
		$T_\mathrm{p}~\mathrm{(K)}$ & $2.79^{+0.79}_{-0.44}\times10^{5}$ & $2.78^{+0.94}_{-0.46}\times10^{5}$\\
		$F_\mathrm{tdediscspec}~\mathrm{(erg~cm^{-2}~s^{-1})}$ & $1.15^{+0.22}_{-0.24} \times 10^{-13}$ & $1.15^{+0.22}_{-0.24} \times 10^{-13}$\\
		$\Gamma$ & $2.07\pm0.26$ & $2.07^{+0.27}_{-0.26}$\\
        $f_\mathrm{sc}$ & - & $3.06^{+4.52}_{-2.61} \times 10^{-3}$\\
        $F_\mathrm{pl}~\mathrm{(erg~cm^{-2}~s^{-1})}$ & $4.79^{+0.65}_{-0.69} \times 10^{-14}$ & $4.73^{+3.19}_{-3.38} \times 10^{-14}$\\
        \hline
	\end{tabular}
\end{table}

\section{SUPPLEMENTARY DATA and TABLES}
We list the log of optical spectroscopy in Table~\ref{tabc1}. The data used to produce Fig.~\ref{fig1} are publicly available on Zenodo ({\href{ https://zenodo.org/records/21335066}{https://doi.org/10.5281/zenodo.21335066}}). 
\begin{table}
	\centering
	\caption{Log of optical spectroscopy.}
	\label{tabc1}
	\begin{tabular}{ccccccc} 
		\hline
        Observation date (UTC) & Phase (d) & Telescope & Instrument & Grism & Exposure (s) \\
        \hline
        2025-01-03 06:28:25 & 29 & NOT & ALFOSC & Grism \#4@1.0"  &   900 \\
        2025-01-17 05:55:25 & 43 & LT & SPRAT &  G4@1.8" & 3$\times$500   \\
        2025-01-19 19:12:34 & 45 & FTN & FLOYDS & red/blu@2.0" & 3600 \\
        2025-01-24 06:05:52 & 50 & NOT & ALFOSC & Grism \#4@0.5" & 2$\times$1650  \\
        2025-01-28 14:15:13 & 54 & FTN & FLOYDS & red/blu@2.0" & 3600 \\
        2025-03-01 14:47:42 & 86 & FTN & FLOYDS & red/blu@2.0" & 3600 \\
        2025-03-29 12:16:16 & 114 & FTN & FLOYDS & red/blu@2.0" & 3600 \\
        2025-04-06 17:36:17 & 121 & FTS & FLOYDS & red/blu@2.0" & 3600 \\
        2025-04-15 06:28:50 & 131 & NTT & EFOSC2 & G13@1.0" & 3$\times$2700  \\
        2025-05-02 13:16:07 & 148 & FTN & FLOYDS & red/blu@2.0" & 3600 \\
        2025-05-09 12:59:30 & 155 & FTN & FLOYDS & red/blu@2.0" & 3600 \\
        2025-05-16 12:33:05 & 162 & FTN & FLOYDS & red/blu@2.0" & 3600 \\
        2025-05-28 11:44:45 & 174 & FTN & FLOYDS & red/blu@2.0" & 3600 \\
        2025-05-30 03:33:12 & 176 & NTT & EFOSC2 &  G13@1.0"  &  3$\times$2700\\
        2025-06-29 22:54:24 & 206 & GTC & OSIRIS+ &  R1000B@1.0"  & 3$\times$900 \\
        2025-08-11 23:31:07 & 249 & NTT & EFOSC2 &  G13@1.0"  &3300  \\
		\hline
	\end{tabular}
\end{table}

\bibliographystyle{raa}
\bibliography{bibtex}

\label{lastpage}

\end{document}